\documentclass{aastex62}

\submitjournal{AJ}

\shorttitle{Radial velocities in outermost disk}
\shortauthors{L\'opez-Corredoira et al.}

\begin{document}

\title{Radial velocities in the outermost disk toward the anticenter}

\author{M.  L\'opez-Corredoira}
\email{martinlc@iac.es}
\affiliation{Instituto de Astrofisica de Canarias, E-38205 La Laguna, Tenerife, Spain}
\affiliation{Departamento de Astrofisica, Universidad de La Laguna, E-38206 La Laguna, Tenerife, Spain}

\author{F. Sylos Labini}
\affiliation{Centro Fermi - Museo Storico della Fisica e Centro Studi e Ricerche Enrico Fermi, I-00186 Rome, Italy}
\affiliation{Istituto dei Sistemi Complessi Consiglio Nazionale delle Ricerche, Via dei Taurini 19, I-00185 Rome, Italy}
\affiliation{INFN Unit Rome 1, Dipartimento di Fisica, Universit\'a di Roma ``Sapienza'', Piazzale Aldo Moro 2, I-00185 Roma, Italy}

\author{P. M. W. Kalberla}
\affiliation{Argelander-Institut f\"ur Astronomie, Auf dem H\"ugel 71, D-53121, Bonn, Germany}

\author{C. Allende Prieto}
\affiliation{Instituto de Astrofisica de Canarias, E-38205 La Laguna, Tenerife, Spain}
\affiliation{Departamento de Astrofisica, Universidad de La Laguna, E-38206 La Laguna, Tenerife, Spain}

\begin{abstract}
We measure the mean Galactocentric radial component of the velocity of stars ($v_R$) in the disk at 8 kpc$<R<28$ kpc in the direction of the anticenter. For this, we use the Apache Point Galactic Evolution Experiment (APOGEE). Furthermore, we compare the result with HI maps along the same line of sight.
We find an increase in positive (expansion) $v_R$ at $R\approx 9-13$ kpc, reaching a maximum of 
$\approx 6$ km/s, and a decrease at large values of $R$ reaching 
a negative (contraction) value of $\approx -10$ km/s for $R>17$ kpc. 
Negative velocities are also observed in 21 cm HI maps,
possibly dominated by local gas emission.
Among the possible dynamical causes for these non-zero $v_R$,
factors such as the effect of the Galactic bar, streams, or mergers do not seem appropriate to explain our observations. An explanation might be the 
gravitational attraction of overdensities in a spiral arm. As a matter of fact, we see a change of regime from positive to negative velocities around  $R\approx 15$ kpc,
in the position where we cross the Outer spiral arm in the anticenter.
The mass in spiral arms necessary to produce these velocities would be about 3\% of the
mass of the disk, consistent with our knowledge of the spiral arms.
Another scenario that we explore is a simple class of out-of-equilibrium systems
 in which radial motions are generally created by the monolithic collapse of isolated self-gravitating overdensities.
\end{abstract}


\section{Introduction}

Disks in spiral galaxies are usually considered to be in a quasi-equilibrium configuration, where stars and other emitters move on steady circular orbits around the center of the galaxy with almost constant velocity. However, most disks exhibit a wealth of non-axisymmetric structures
(Rix \& Zaritsky 1995; Laine et al. 2014); 
about one third of them are substantially lopsided at $>$2.5 times the scale length of the disk, 
although the spiral pattern couples significantly to the estimate of the intrinsic ellipticity and their measurement may represent an upper limit to the true potential triaxiality. Lopsidedness is quite typical in disk galaxies and this may be interpreted as a pattern of elliptical orbits (Baldwin et al. 1980; Song et al. 1983). Non-circular streaming motions have also been observed in the gas motions of other galaxies (Trachternach et al. 2008; Sellwood \& Zanmar S\'anchez 2010), but the stellar kinematics is usually more regular and symmetric than the gas kinematics (Pizzella et al. 2008). A new analysis using tracers of the stellar populations should be investigated. This can be more accurately done in the Milky Way, which is the goal of the present paper.

In our Galaxy, Siebert et al. (2011) and Williams et al. (2013) measured  a significant 
average gradient in Galactocentric radial velocity ($\langle v_R\rangle $) outwards within 6 kpc$\lesssim R\lesssim 9$ kpc. Carrillo et al. (2018) reached $R=10$ kpc with Gaia-DR1 data. L\'opez-Corredoira \& Gonz\'alez-Fern\'andez (2016; hereafter LG16) and Tian et al. (2017) measured the mean radial Galactocentric velocity in the anticenter direction, reaching distances up to $R=16$ kpc using red clump giants as standard candles, and they obtained significant positive values of $\lesssim +10$ km/s. Wang et al. (2018b) confirmed these results up to $R\approx 13$ kpc with K giant stars. 
Gaia Collaboration (2018, hereafter G18) provided a full kinematic information in all lines of sight for different populations with the recent release of Gaia-DR2 data that include
parallaxes, proper motions, and radial velocities, but also are limited to $R\lesssim 13$ kpc; this range
was extended up to $R\approx 20$ by L\'opez-Corredoira \& Sylos Labini (2018, hereafter LS18) with a method to deconvolve parallax errors that allows the use of data with large errors in distance.
G18 or Antoja et al. (2018) or LS18 could show that the Galactic disk as a time-independent axisymmetric component is definitively an incorrect image of our Galaxy.
Harris et al. (2018) took A and F stars as standard candles and also tried to measure the radial velocities up to distances of 15 kpc, but the uncertainties of their average values of several km/s are too large to measure small values of $\langle v_R\rangle $. Using star counts that fit the different components of the Galaxy
(L\'opez-Corredoira \& Molg\'o 2014) or doing analyses of metallicity distributions in spectroscopic surveys (L\'opez-Corredoira et al. 2018), it was known that the outer disk extends up to Galactocentric distances much farther away than 20 kpc, so one might explore the kinematics in this very outer region, possibly dominated by non-equilibrium
dynamics. In this paper, we aim to obtain measurements of the radial velocity up to these greater Galactocentric distances, using the stellar spectrophotometric parallaxes derived by several methods by the ``Apache Point Galactic Evolution Experiment'' (APOGEE). We give information about the data source and distance derivation in \S \ref{.data}. The measurements of the mean radial velocity of these stars in the anticenter direction are given in \S \ref{.radvel}, where we also include the
comparison of HI gas in that direction.

These measurements may be interpreted in terms of ellipticity, net migration of stars, associated with causes
such as streams, bars, spiral arms, mergers, or non-equilibrium states derived from the particular evolution of isolated disks. All of these kinematical and dynamical analyses
are carried out in \S \ref{.interp}. Particular attention is paid to a scenario that has been proposed in a recent work by Benhaiem
et al. (2017; hereafter B17), where it was shown that a far-from-equilibrium dynamics may give rise to a disk with 
 transient  spiral-like arms  in a manner completely different to that envisaged by the perturbative mechanisms for generating such structure: the transient nature of the outer part of the disk and of the arms is related to the presence of 
coherent radial velocities, with both negative (i.e., contraction)  and positive  (i.e., expansion) signs.
On the other hand, we show that galaxy-like structures formed by magnetohydrodynamical 
simulations (e.g., Grand et al. 2018) typically form axisymmetric structures without coherent radial velocities.
Finally, we summarize our results and interpretation in a concluding section (\S \ref{.concl}).

\section{Data}
\label{.data}

APOGEE (Majewski et al. 2017) is one of the surveys included in the 
Sloan Digital Sky Survey (SDSS) III
(Eisenstein et al. 2011) and IV (Blanton et al. 2017). APOGEE employs a 
pair of custom-built
  high-resolution (R$\equiv \lambda/\delta\lambda \sim 22,500$) 
multi-object
$H-$band (1.5--1.7 $\mu$m) spectrographs (Wilson et al. 2010, 2012)
on 2.5-metre telescopes with a 3$^\circ $ FOV: the SDSS telescope at
Apache Point Observatory (New Mexico, US; see Gunn et al. 2006) and the 
du Pont telescope at Las Campanas (Chile).
The project is collecting hundreds of thousands of stellar spectra with 
high signal-to-noise ratios across the Milky Way, focusing on the regions 
where dust
causes dramatic extinction at optical wavelengths, namely the disk and 
the central parts of the Galaxy.
The APOGEE observations are providing a chemical map of our Galaxy with 
unprecedented quality
(Badenes et al. 2018; Fern{\'a}ndez-Alvar et al. 2018;
Fu et al. 2018; Garc\'\i a-Dias et al. 2018; Garc\i{\i}a-P\'erez et al. 
2018; Hayes et al. 2018;
L\'opez-Corredoira et al. 2018; Mackereth et al. 2019; Palicio et al. 2018;
Souto et al. 2018; Weinberg et al. 2018; Wilson et al. 2018).

The 14th data release of the SDSS (DR14; Abolfathi et al. 2018; Holtzman 
et al. 2018; J\"onsson et al. 2018)
includes millions of APOGEE spectra for approximately 263,000 stars 
observed from
Apache Point Observatory, which have been processed by the project's 
pipelines
(Nidever et al. 2015; Garc\'\i a-P\'erez et al. 2016)
to derive radial velocities, atmospheric parameters, and abundances for 
more than
20 chemical elements, typically with a precision better than 0.1 dex.
Distances to the stars have been estimated by four different
methods and included in a value-added catalog released in conjunction 
with DR14:
the StarHorse code (Santiago et al. 2016),
the code described in Wang et al. (2016), the isochrone-matching technique
described in Schultheis et al. (2014), and the distance code
of Holtzman et al. (2018). The agreement among the four codes is fair, 
typically
within 20\%. We have adopted for our analysis the average values of the 
available
estimates, so the error of this average distance is expected to be $\sim 
$10\%.

\begin{figure}
\vspace{1.5cm}
\centering
\includegraphics[width=8cm]{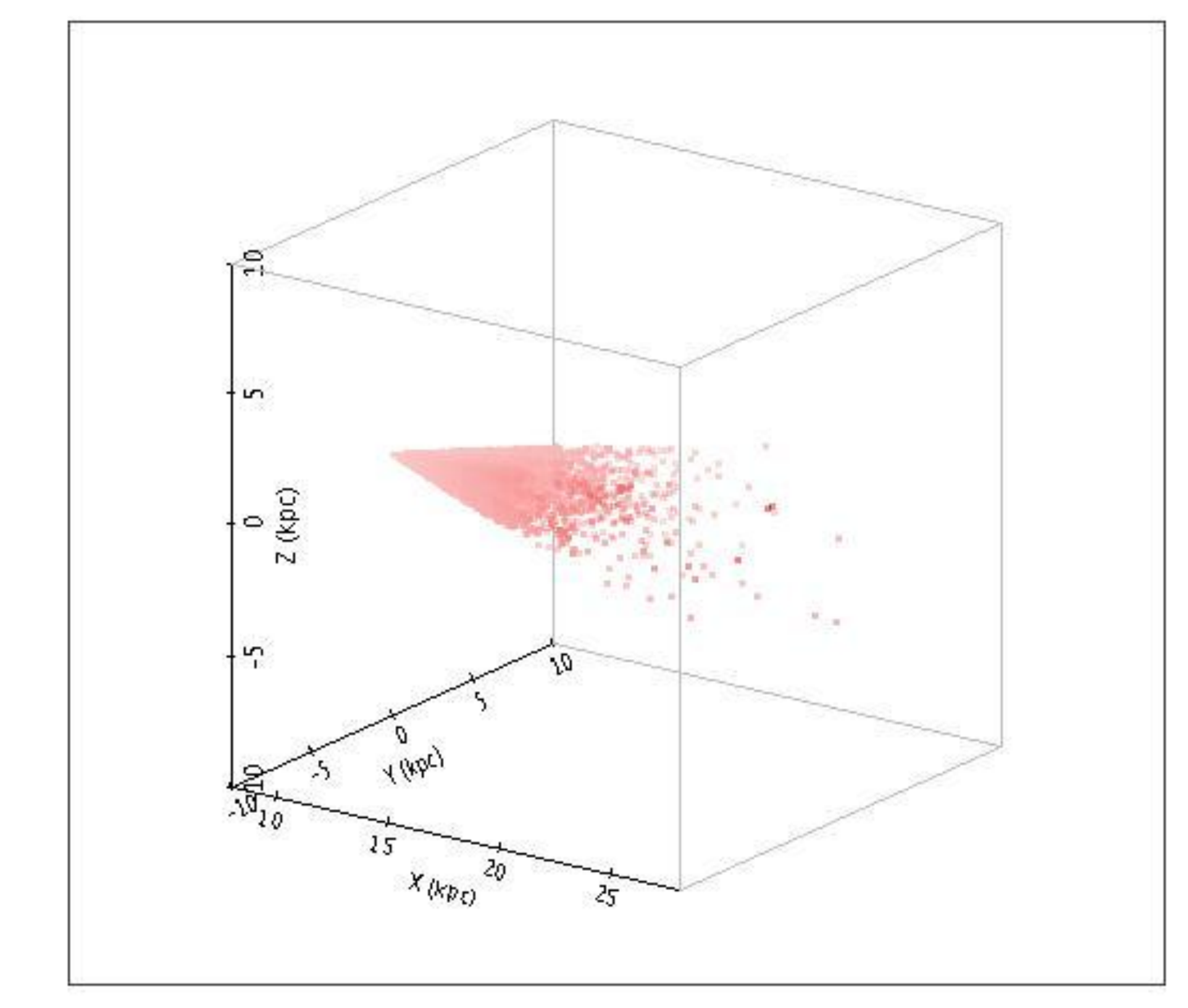}
\caption{Three-dimensional distribution in Cartesian Galactocentric coordinates 
of the APOGEE sources in the anticenter subsample with
$|\ell-180^\circ |<20^\circ $, $|b|<10^\circ$.}
\label{Fig:3Danticenter} 
\end{figure}

The spatial distribution of the stars considered here spans a range of Galactic 
longitudes accessible from observatories in Northern-hemisphere.
The range of Galactocentric distances explored in this paper is between $R=6$ and 28 kpc,
always at $|z|<5$ kpc. In Fig. \ref{Fig:3Danticenter} we show the distribution
in cartesian Galactocentric coordinates ($X_\odot =R_\odot $) 
of the 16\,105 sources in the anticenter direction that we will use in this paper:
a decreasing density of sources with increasing $X$ is evident. The number of sources for each value of $R$ is
given in Table \ref{Tab:anticenter}.

The stars in this range are thin- and thick-disk stars; the disk reaches a distance
of 25 kpc from the Galactic center or even farther (L\'opez-Corredoira et al.
2018), with some significant contamination by halo stars only for $R\gtrsim 20$ kpc (L\'opez-Corredoira \& Molg\'o 2014; L\'opez-Corredoira et al. 2018).  
Although some authors attributed some excess stars in the plane at $R\approx 20$
 kpc (the Monoceros Ring) to a putative tidal debris of a dwarf galaxy (Sollima et al. 2011), or
a bending of the disk or a corrugation pattern linked to the interaction with a passing satellite
(G\'omez et al. 2016; Yanny \& Newberg 2016), L\'opez-Corredoira
\& Molg\'o (2014) and Wang et al. (2018a) 
have shown that a flared disk+halo is 
the most likely explanation without the need for any extragalactic element. 
We assume this last position here.

\section{Galactocentric radial velocities}
\label{.radvel}

\subsection{Transformation of Heliocentric into Galactocentric Radial Velocities}

The three components of the velocity derived from proper motions and radial velocities 
were already analyzed for Gaia data (G18; LS18). Here, however, we will only
use the radial velocities. APOGEE provides independent information with a determination of distances different from parallax in Gaia that is more precise for the largest heliocentric distances, so we can reach greater distances with APOGEE than with Gaia. Moreover, given that our analyses will be restricted to the anticenter, proper motions make a negligible contribution to the radial component of the Galactocentric velocity.
With only the radial heliocentric velocity $v_r$, we can obtain
the Galactocentric radial velocity ($v_R$) through (LG16, Eq. 4):
\begin{equation}
\label{VR}
v_R=-\frac{v_r}{\cos (\phi +\ell )\cos b}-\frac{\cos \ell}{\cos (\phi +\ell)}U_\odot -
\frac{\sin \ell}{\cos (\phi+\ell )}V_{g,\odot }
\end{equation}\[
\ \ \ \ \ \ +\frac{\tan b}{\cos (\phi +\ell)}
(W-W_\odot )+\tan (\phi +\ell )V_\phi
,\]
where $(\ell ,b)$ are the Galactic coordinates, $\phi $ is the Galactocentric azimuth (such that $x=R\cos \phi$, $y=R\sin \phi$, $\phi _\odot =0$); ($U_\odot $, $V_\odot$, $W_\odot$) is the velocity of the Sun with respect to the local standard of rest (LSR),
$V_{g,\odot }=V _\phi (R_\odot ,z=0)+V_\odot $; 
$V_\phi (R,z)$ is the azimuthal velocity, and $W(R,z)$ is the vertical motion.
In the calculation of the average or median of $v_R$, we neglect the possible asymmetries
of random peculiar motions.

The disadvantage of Eq. (\ref{VR}) is that it is model-dependent, since we need to know the values of $V_\phi$ and $W$, but we can make an appropriate selection of regions in which the dependence on them is very small. In particular, as we will see next, toward the anticenter, the dependence on these two functions is negligible.

Here, we adopt the values $R_\odot =8$ kpc; $V_{g,\odot }=244\pm 10$ km/s, 
$U_\odot =10\pm 1$ km/s (Bovy et al. 2012), and 
$W_\odot =7.2\pm 0.4$ km/s  (Sch\"onrich et al. 2010).
The azimuthal velocity $V_\phi (R,z=0)=V_{c,0}(R)-V_{a,0}(R)$ where $V_{c,0}$ is the rotation curve, taken from Sofue (2017, Fig. 4), and $V_{a,0}$ is the asymmetric drift at $z=0$.
The contribution from the asymmetric drift $V_{a,0}$ is negligible, especially for the anticenter region;
we include its small contribution as 
\begin{equation}
V_{a,0}(R)=\frac{f_B(R)\sigma _R(R)^2}{V_c(R)} \;, 
\end{equation} 
with $f_B(R)\approx 0.27\,R({\rm kpc})$ (Bovy et al. 2012, Fig. 4/top). $V_{a,0}$
has a value $\sim 10$ km/s for $R\lesssim 15$ kpc and somewhat larger for larger $R$.
The generalized expression with dependence on $z$ that takes into account
the lower rotation speed for $z\ne 0$ is taken as 
$V_\phi (R,z)=V_\phi (R,z=0)-19.2$ km/s$\left(\frac{|z|}{\rm kpc}\right)^{1.25}$
(Bond et al. 2010). As systematic error we take 
$\Delta V_\phi (R,z)=\Delta V_{g,\odot }=10$ km/s 
for $R<20$ kpc and $\Delta V_\phi (R,z)=(R({\rm kpc})-10)$ km/s for $R\ge 20$ kpc,
in agreement with the errors given in Sofue (2017, Fig. 4). 
We set $W=0$, considering negligible the contribution of possible vertical motions; but we take into account the range of possible
values in the calculation of systematic errors using $\Delta W\equiv W_{\rm best\ fit}$ of
Eq. (15) of L\'opez-Corredoira et al. (2014).

\subsection{Analysis in the Anticenter In-plane Region}

\begin{figure*}
\vspace{1.5cm}
\centering
\includegraphics[width=12cm]{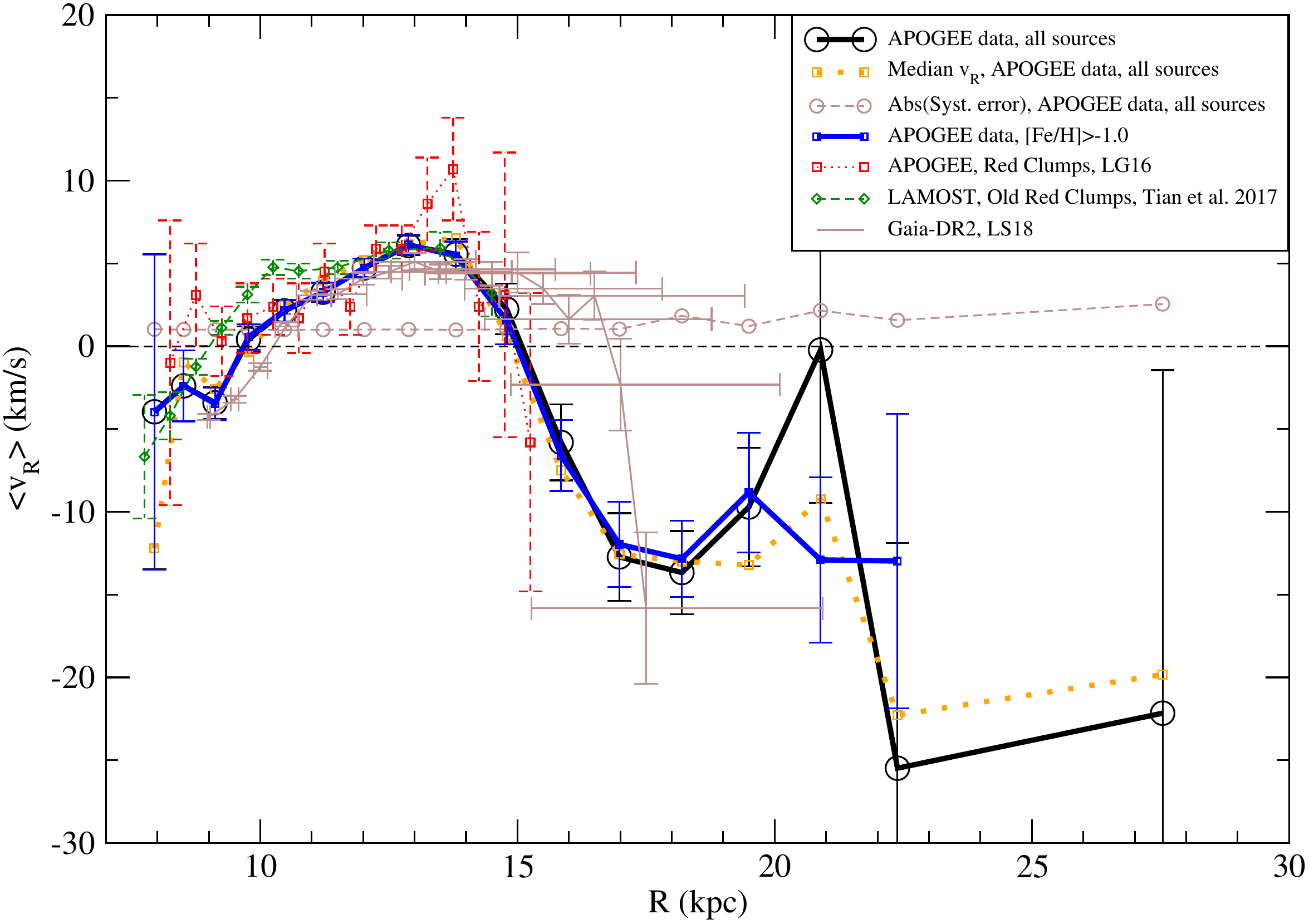}
\caption{Average radial Galactocentric velocity derived from Eq. (\protect{\ref{VR}}) with radial heliocentric velocities from APOGEE within a region close to the Galactic center--Sun line, including either all of the stars or only with stars with [Fe/H]$>$-1. The median of the first sample is also plotted. Other measurements in the literature (LG16, Tian et al. 2017, LS18) are plotted for $R<18$ kpc. Error bars represent statistical errors only. Systematic errors due to uncertainties in the parameters are plotted separately.}
\vspace{.2cm}
\label{Fig:anticenter}
\end{figure*}

First, we choose the lines of sight symmetrically around the anticenter, which minimizes the systematic errors. We choose $|b|<10^\circ$, which makes the contribution of $W$ or $W_\odot $ very low; and $|\ell-180^\circ |<20^\circ $, to make $\cos (\phi+\ell )$ 
close to one for all of the stars and $\langle \tan (\phi+\ell )\rangle$ close to zero, which reduces the impact of the error in $V_\phi$.
The average Galactocentric velocities are shown in Fig. \ref{Fig:anticenter} for $R-R_\odot <20$ kpc, together
with other results of previous studies for $R<18$ kpc in fair agreement with our data.
Note that the selected regions in the different surveys are not exactly the same,
although all are centered in the anticenter region,
so no perfect agreement in the results is expected.
The median of $V_R(R)$ is also plotted for the first sample, with results very similar
to the average; thus, we will use the ``average'' hereafter. 
In Table \ref{Fig:anticenter} we give the numerical values. Bins are divided into
$\Delta \log R({\rm kpc})=0.03$, so as to have $N\ge 6$ stars in each of them.
We plot the average velocity for all the stars, and separately for those at 
[Fe/H]$>-1$. The second subsample, which is restricted in metallicity, is almost pure disk stars with a
very small contamination of the halo population (L\'opez-Corredoira et al. 2018). Both of them show
compatible results for $\langle V_R\rangle$ because, even for the first sample, disk stars
 dominate. The outcome is an increase in positive velocity (expansion of the disk) from 
$R\approx 9$ kpc to $R\approx 13$ kpc, reaching a maximum of $\approx 6$ km/s, and a decrease outwards, reaching $\approx -10$ kpc for $R>17$ kpc (negative means contraction of the disk). The signal is significantly larger than the statistical errors (dispersion in the measurements of $v_r$) and systematic errors (due to the uncertainties in the parameters used in Eq. (\ref{VR})).

\begin{figure}
\vspace{1.5cm}
\centering
\includegraphics[width=8cm]{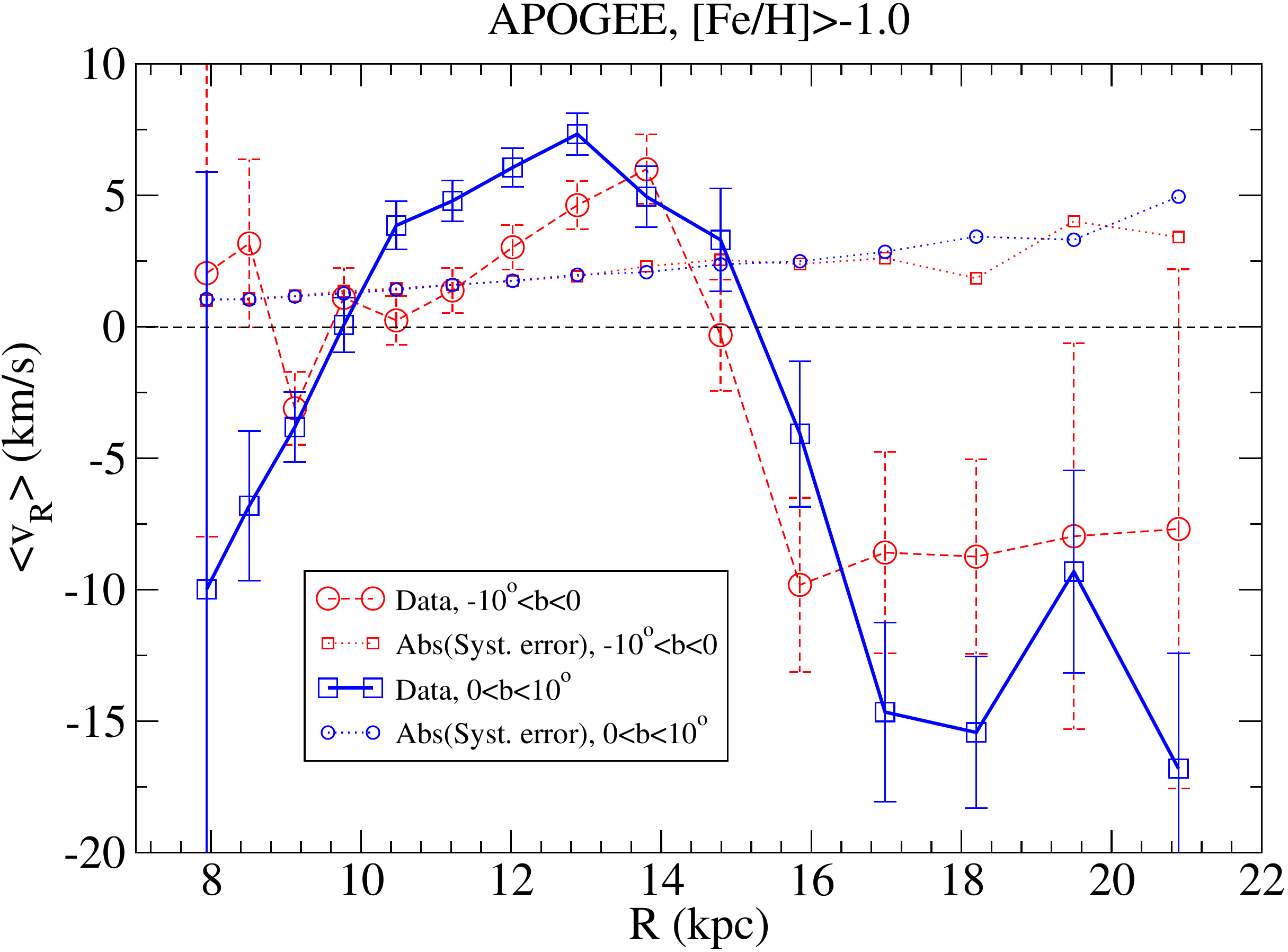}
\caption{Average radial Galactocentric velocity derived from Eq. (\protect{\ref{VR}}) with radial heliocentric velocities from APOGEE within a region close to the Galactic center--Sun line, including only stars with [Fe/H]$>$-1, separating northern and southern Galactic data. Error bars represent statistical errors only. Systematic errors due to uncertainties in the parameters are plotted separately.}
\vspace{.2cm}
\label{Fig:vr2}
\end{figure}

\begin{figure}
\vspace{1.5cm}
\centering
\includegraphics[width=8cm]{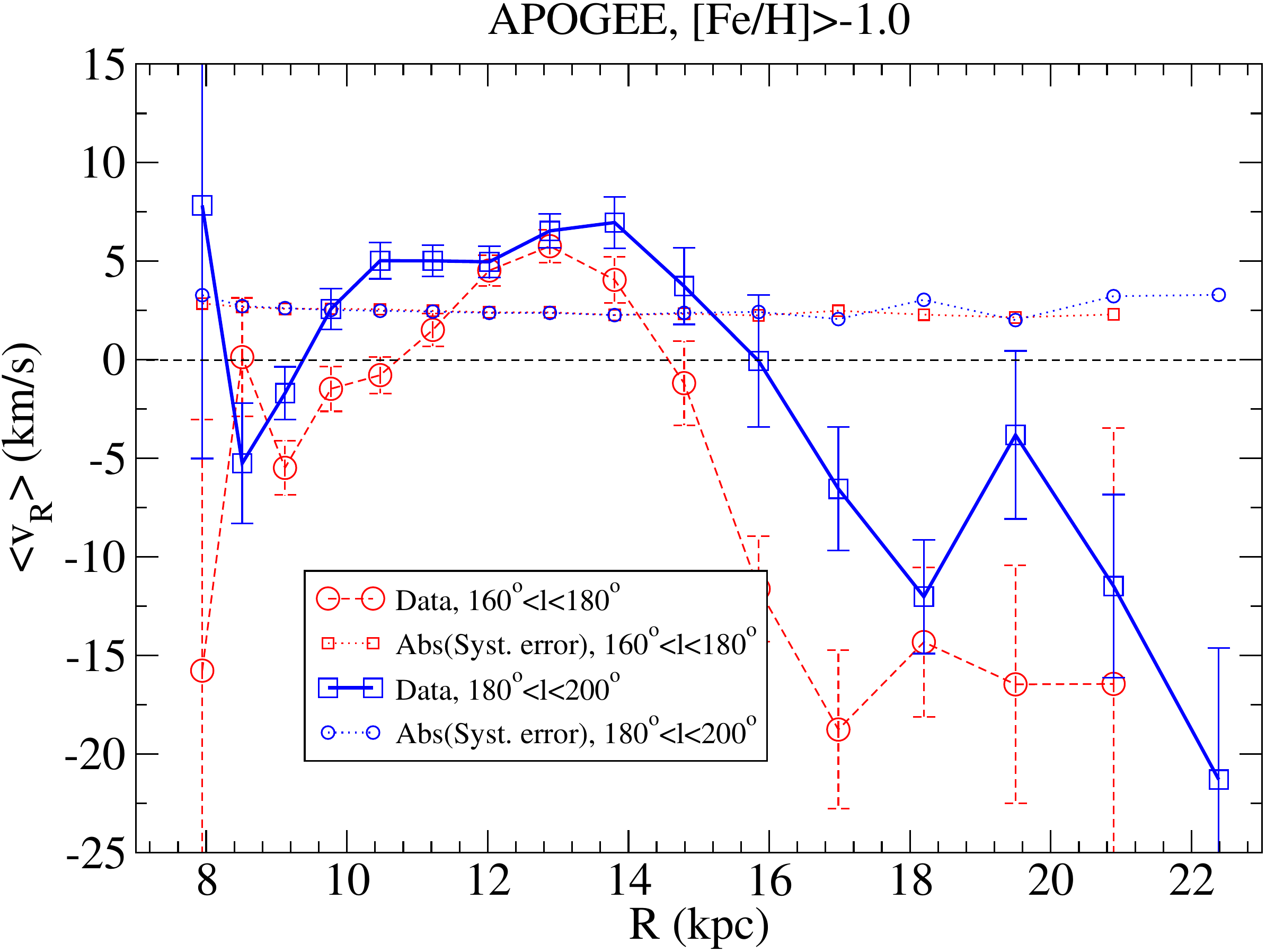}
\caption{Average radial Galactocentric velocity derived from Eq. (\protect{\ref{VR}}) with radial heliocentric velocities from APOGEE within a region close to the Galactic center--Sun line, including only stars with [Fe/H]$>$-1, separating $\ell <180^\circ$  and $\ell >180^\circ $ data. Error bars represent statistical errors only. Systematic errors due to uncertainties in the parameters are plotted separately.}
\vspace{.2cm}
\label{Fig:vr3}
\end{figure}

If we divide the sample with [Fe/H]$>-1$ into two subsamples corresponding to the northern and southern Galactic hemispheres, the results show some differences (Fig. \ref{Fig:vr2}) but within similar trends.
The systematic errors are larger due to the $\Delta W$ term when $\langle z\rangle$ is significantly different from zero. If the division is carried out into $\ell <180^\circ $ and $\ell >180^\circ $, however, the results are significantly different (Fig. \ref{Fig:vr3}), indicating also that there is an important dependence of the radial velocity on the azimuth.

\begin{figure}
\vspace{1.5cm}
\centering
\includegraphics[width=8cm]{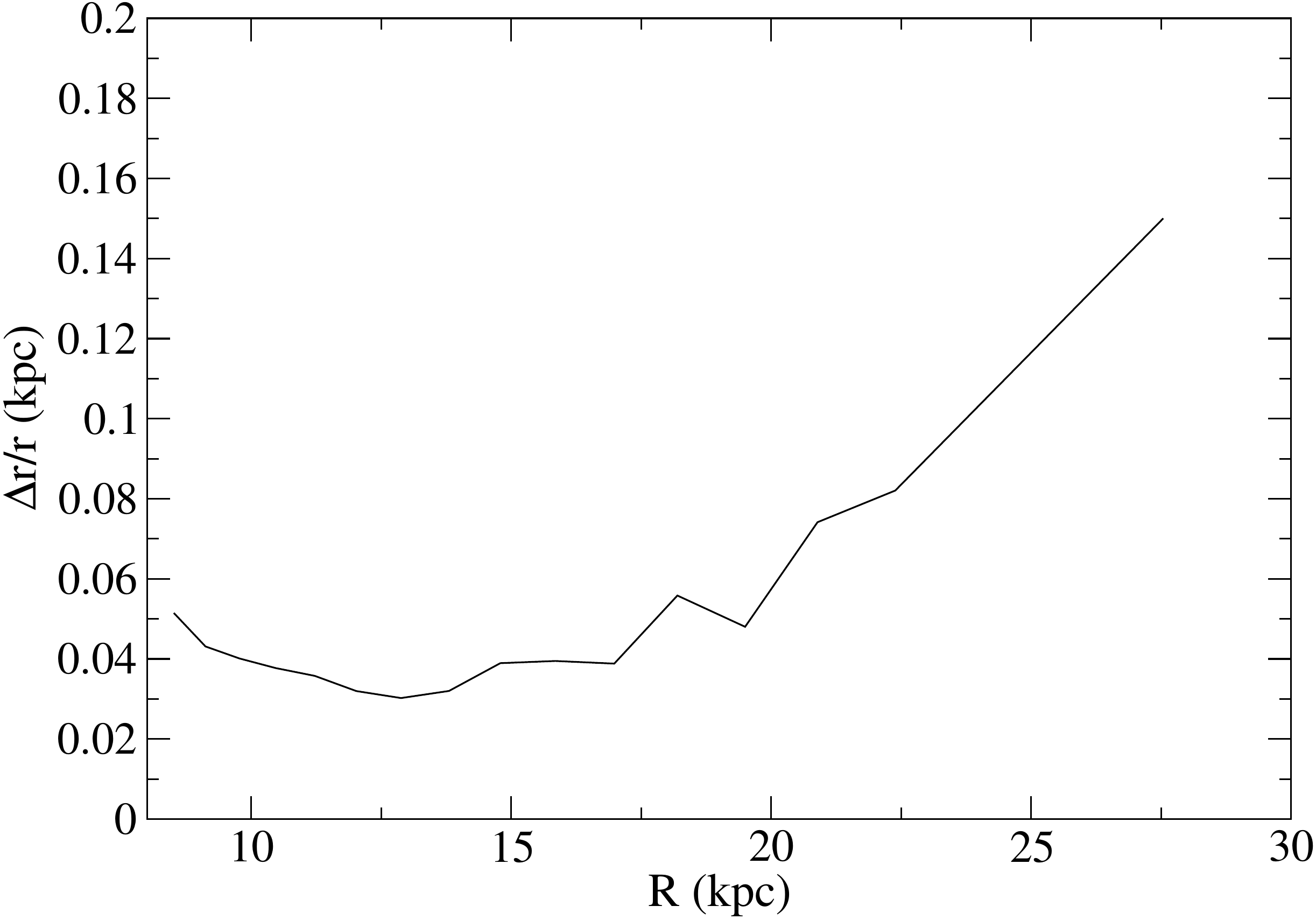}
\caption{Average relative error of heliocentric distance in the APOGEE sources as a function of Galactocentric distance.}
\vspace{.2cm}
\label{Fig:errordist}
\end{figure}

The errors in the determination of heliocentric distance for any given star (around 10\% as stated;  Fig. \ref{Fig:errordist} shows the average error as a function of Galactocentric distance for this line of sight) might produce a smoothing in the function $\langle v_R (R) \rangle $, but, given that the width of the bins is of the order of these errors, the variation of the function should not be important. The positive or negative signal cannot be due to errors in distance estimates.

\begin{table*}
\caption{Average radial Galactocentric velocity derived from Eq. (\protect{\ref{VR}}) with radial heliocentric velocities from APOGEE within a region close to the Galactic Center--Sun line, including All of the stars or only with stars with [Fe/H]$>$-1. Bins have $N\ge 6$ and are separated by $\Delta \log R({\rm kpc})=0.03$. The corresponding numbers of stars in each bin are indicated in the columns 4, 5; the r.m.s. of the radial velocities is
given in columns 6, 7.}
\begin{center}
\begin{tabular}{ccccccc}
$R$ (kpc) & $\langle v_R\rangle _{\rm all}$ (km/s) & $\langle v_R\rangle _{{\rm [Fe/H]}>-1}$ (km/s) & $N_{\rm all}$  & $N_{{\rm [Fe/H]}>-1}$ & $\sigma_{\rm all}$ (km/s)  & $\sigma_{{\rm [Fe/H]}>-1}$ (km/s) \\ \hline
  7.9 & -4.0$\pm $ 9.5(stat.)$\pm $ 1.0(syst.) & -4.0$\pm $ 9.5(stat.)$\pm $ 1.0(syst.) & 16 & 16 & 38.1 & 39.0 \\
  8.5 & -2.4$\pm $ 2.1(stat.)$\pm $ 1.0(syst.) & -2.4$\pm $ 2.1(stat.)$\pm $ 1.0(syst.) & 289 & 289 & 36.5 & 36.5 \\
  9.1 & -3.4$\pm $ 1.0(stat.)$\pm $ 1.0(syst.) & -3.5$\pm $ 1.0(stat.)$\pm $ 1.0(syst.) & 1246 & 1244 & 33.9 & 34.0 \\
  9.8 &  0.4$\pm $ 0.8(stat.)$\pm $ 1.0(syst.) & 0.6$\pm $ 0.8(stat.)$\pm $ 1.0(syst.) & 2011 & 2007 & 34.6 & 34.7 \\
  10.5 &  2.1$\pm $ 0.7(stat.)$\pm $ 1.0(syst.) & 2.2$\pm $ 0.7(stat.)$\pm $ 1.0(syst.) & 2496 & 2493 & 32.7 & 33.0 \\
  11.2 &  3.3$\pm $ 0.6(stat.)$\pm $ 1.0(syst.) & 3.3$\pm $ 0.6(stat.)$\pm $ 1.0(syst.) & 2979 & 2979 & 31.4 & 31.5 \\
  12.0 &  4.7$\pm $ 0.6(stat.)$\pm $ 1.0(syst.) & 4.7$\pm $ 0.6(stat.)$\pm $ 1.0(syst.) & 3000 & 2998 & 30.5 & 30.5 \\
  12.9 &  6.1$\pm $ 0.6(stat.)$\pm $ 1.0(syst.) & 6.1$\pm $ 0.6(stat.)$\pm $ 1.0(syst.) & 2022 & 2021 & 27.1 & 27.1\\
  13.8 &  5.6$\pm $ 0.9(stat.)$\pm $ 1.0(syst.) & 5.4$\pm $ 0.9(stat.)$\pm $ 1.0(syst.) & 1099 & 1094 & 30.2 & 28.9 \\
  14.8 &  2.3$\pm $ 1.5(stat.)$\pm $ 1.0(syst.) & 1.6$\pm $ 1.4(stat.)$\pm $ 1.0(syst.) & 451 & 449 & 32.6 & 30.7 \\
  15.8 & -5.8$\pm $ 2.3(stat.)$\pm $ 1.1(syst.) & -6.6$\pm $ 2.1(stat.)$\pm $ 1.0(syst.) & 196 & 191 & 32.1 & 30.0 \\
  17.0 & -12.7$\pm $ 2.6(stat.)$\pm $ 1.0(syst.) & -12.0$\pm $ 2.6(stat.)$\pm $ 1.1(syst.) & 109 & 106 & 27.6 & 26.6 \\
  18.2 & -13.7$\pm $ 2.5(stat.)$\pm $ 1.8(syst.) & -13.0$\pm $ 2.3(stat.)$\pm $ 1.8(syst.) & 87 & 85 & 23.4 & 21.3 \\
  19.5 & -9.7$\pm $  3.6(stat.)$\pm $  1.2(syst.) & -8.8$\pm $ 3.6(stat.)$\pm $ 1.2(syst.) & 54 & 53 & 26.4 & 26.7 \\
  20.9 & -0.2$\pm $  9.2(stat.)$\pm $  2.2(syst.) & -12.9$\pm $ 5.0(stat.)$\pm $ 1.9(syst.) & 32 & 28 & 52.2 & 26.5 \\
  22.4 & -25.5$\pm $ 13.6(stat.)$\pm $  1.6(syst.) & -13.0$\pm $ 8.9(stat.)$\pm $ 2.2(syst.) & 9 & 8 & 40.8 & 24.8 \\
  23.9 & -- & -- & 0 & 0 & -- & -- \\
 25.7 & -- & -- & 1 & 1 & -- & -- \\
  27.5 & -22.2$\pm $ 20.7(stat.)$\pm $ 2.5(syst.) & -- & 6 & 3 & 50.7 & -- \\ \hline
\label{Tab:anticenter}
\end{tabular}
\end{center}
\end{table*}


\subsection{Radial velocity from the 21 cm line in the anticenter}

We wonder whether this feature of non-zero Galactocentric radial velocity can also be observed in neutral hydrogen. The anticenter is a region that is usually avoided for the analysis of the 21 cm line, 
because distances should be derived from heliocentric velocities assuming circular orbits, and thus one would expect zero velocity (except for the corrections for the motion of the Sun with respect to the LSR) at any distance, so there is no way to disentangle the integrated flux. Nonetheless, there might be some net redshift/blueshift or some distortion that could make us suspect that the orbits are not perfectly circular in the gas in the outer disk. 

A simple exercise can illustrate this. Assuming that gas and stars
have a similar mean velocity, the integrated flux of the antenna temperature
in HI in the anticenter should be (corrected to Galactocentric coordinates)
\begin{equation}
\label{radio}
T(v)\propto \int _{R_\odot }^\infty  
\rho _{HI}(R)\exp\left(-\frac{(v-v_R(R))^2}{2\sigma _R(R)^2}\right)dR
,\end{equation}
where $\rho _{HI}(R)$ is the HI density at distance $R$ and $\sigma _R$ is
the dispersion of $v_R$. We 
assume a constant velocity dispersion for the HI gas. The
observational evidence for a constant dispersion is indirect.
Dickey et al. (2009) have shown that in the outer part of the
Milky Way up to the observational limit at radii of $\sim $25 kpc the ratio of
the emission to the absorption, which determines the mean spin
temperature of the gas, stays nearly constant. This implies on average a
constant mixture between cold and warm neutral medium, hence a constant
velocity dispersion.

The state of the HI is determined by gas-phase reactions, in equilibrium
situations with isotropic velocity dispersions for individual gas
clouds. The velocity dispersion of the gas is not related to the
dynamical state of the stellar population, but depends on the star
formation rate (SFR). Dib et al. (2006) investigated the
relationship between the velocity dispersion of the gas, the supernova
rate, and feedback efficiency with three-dimensional numerical
simulations of supernovae-driven turbulence in the interstellar medium and
explained the approximate constancy of observed velocity dispersion
profiles in the outer parts of galactic disks. Thermal instabilities in
the HI gas, driven by supernovae, determine spin temperatures,
composition, and velocity dispersion of the HI.

Tamburro et al. (2009) studied a sample of galaxies with a resolved
systematic radial decline in the HI line width, implying a radial
decline in kinetic energy density of HI. They find a correlation between
the kinetic energy of gas and SFR with a slope close to unity and a
similar proportionality constant in all objects.

Observations of velocity dispersions in external galaxies are faced with
difficulties in disentangling the radial HI surface density, the
rotation curve, and the HI velocity dispersion.
O'Brien et al. (2010) observed the HI in edge-on galaxies, where
it is in principle possible to measure the force field in the halo
vertically and radially from gas layer flaring and rotation curve
decomposition respectively. They found (their Fig. 21) radial
fluctuations of the velocity dispersions, but on average the dispersions
are constant in the outer parts of the disks. In all cases a distinct
flaring is observed (O'Brien et al. 2010, Fig. 25). 
In fact, for a constant velocity dispersion the gradient of the force field in the
vertical direction implies exponential flaring for the gas layer. Such a
dependence is also observed in the Milky Way (Kalberla et al. 2007).

\begin{figure}
\vspace{1.5cm}
\centering
\includegraphics[width=8cm]{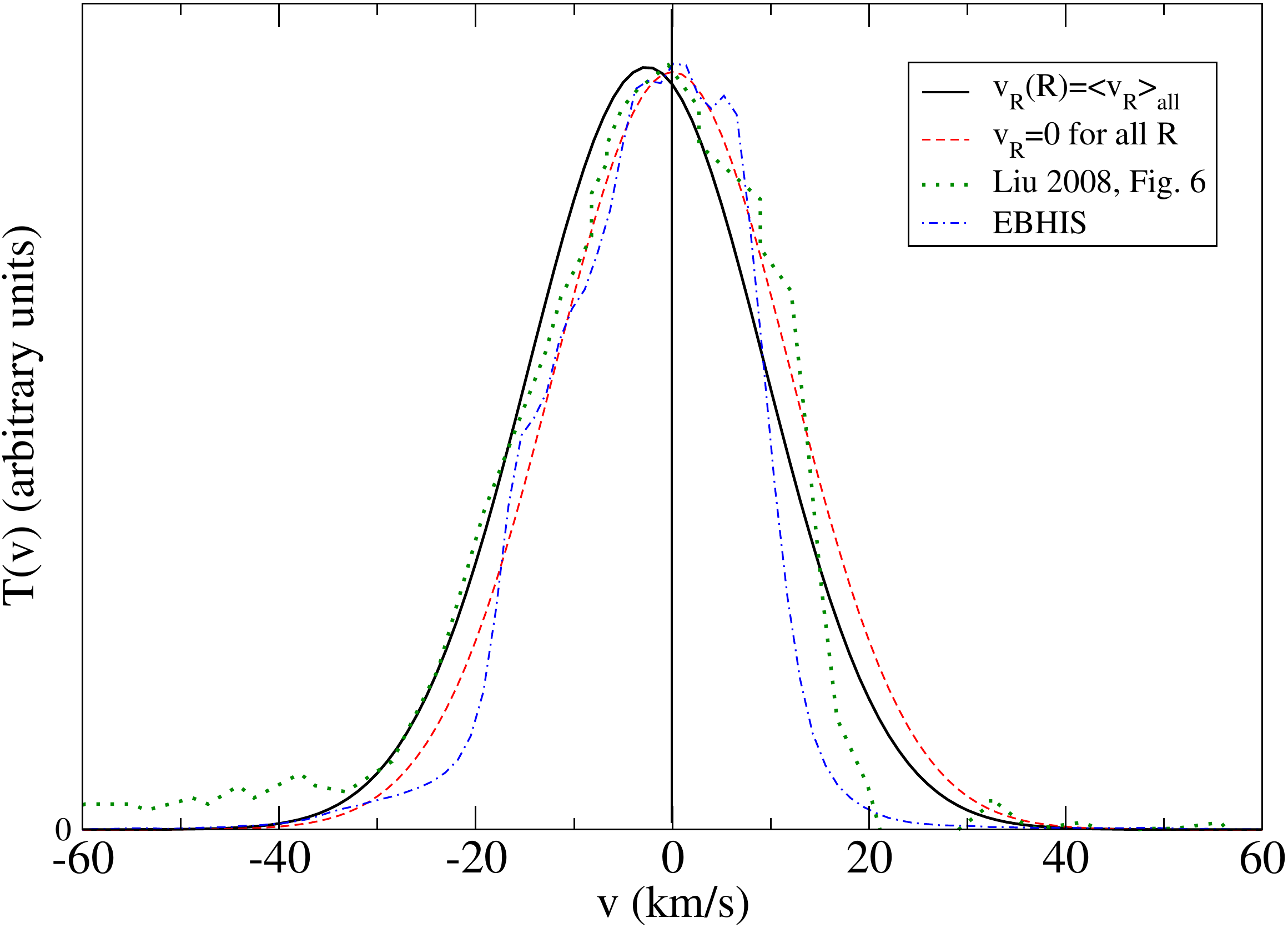}
\caption{Expected profile of the  HI line in observations towards the anticenter
for $\sigma =12$ km/s, gas scale-length of 3.75 kpc and mean radial velocities
null or equal to the observed APOGEE data for all sources (2nd. column in Table
\ref{Tab:anticenter}). 
Also plots of HI observation profiles
of Liu (2008, Fig. 6) with a FWHM beam size of 6$^\circ $ (green, dotted) 
and EBHIS survey (Winkel et al. 2016) with a narrow beam of FWHM=$0.2^\circ $
(blue, dash-dot) are included.}
\vspace{.2cm}
\label{Fig:calc_HI}
\end{figure}

In Fig. \ref{Fig:calc_HI} we plot the expected profile of the HI line in observations 
toward the anticenter for $\sigma _R(R)=12$ km/s, $\rho _{HI}(R)\propto \exp(-R/h_{R,HI})$ 
with a gas scale length of $h_{R,HI}=3.75$ kpc (Kalberla \& Dedes 2008), and mean radial velocities
null or equal to the observed APOGEE data for all sources (second column in Table
\ref{Tab:anticenter}).
The result shows a small asymmetry with respect to perfect 
circular orbits with slightly larger dispersion for negative velocities: $T(-v)>T(v)$.
The effect is not very large, but it might be significant. 
Note, however, that, with these parameters, 40\% of the temperature comes from 
gas at $R<10$ kpc and 80\% of the temperature from gas at $R<15$ kpc, so the outermost disk
features are not dominant in the maps of gas flux.

As a matter of fact, the results of Liu (2008, Fig. 6) 
or those from the EBHIS survey (Winkel et al. 2016)\footnote{Data provided at 
https://www.astro.uni-bonn.de/hisurvey/AllSky$\_$profiles/index.php. See also
HI4PI Collaboration (2016).}, reproduced in our Fig. \ref{Fig:calc_HI}, show an even larger asymmetry, 
which would confirm this trend; the larger asymmetry might be due
to a $h_{R,HI}>3.75$ kpc or to higher amplitude of negative $v_R$ for $R<10$ kpc
(which is affected by very large errors in our APOGEE sample) 
or because the stars in our region ($160^\circ <\ell <200^\circ $,
$|b|<10^\circ $) have on average lower radial velocities than stars in the
region of the anticenter alone, or simply because gas presents higher 
asymmetries than the stars in the velocity distribution, as observed in other
galaxies (Pizzella et al. 2008).

The velocity distribution of the 21 cm line emission is usually affected
by turbulent motions (Kalberla \& Kerp 2009). In Fig. \ref{Fig:calc_HI}, we compare
observations with significantly different FWHM beam sizes of $\sim 6^\circ $ and $\sim 0.2^\circ$. It appears implausible that the asymmetries in both cases are caused by random turbulent motions, but we cannot exclude this possibility.

\begin{figure*}
\vspace{1.5cm}
\centering
\includegraphics[width=10cm,angle=-90]{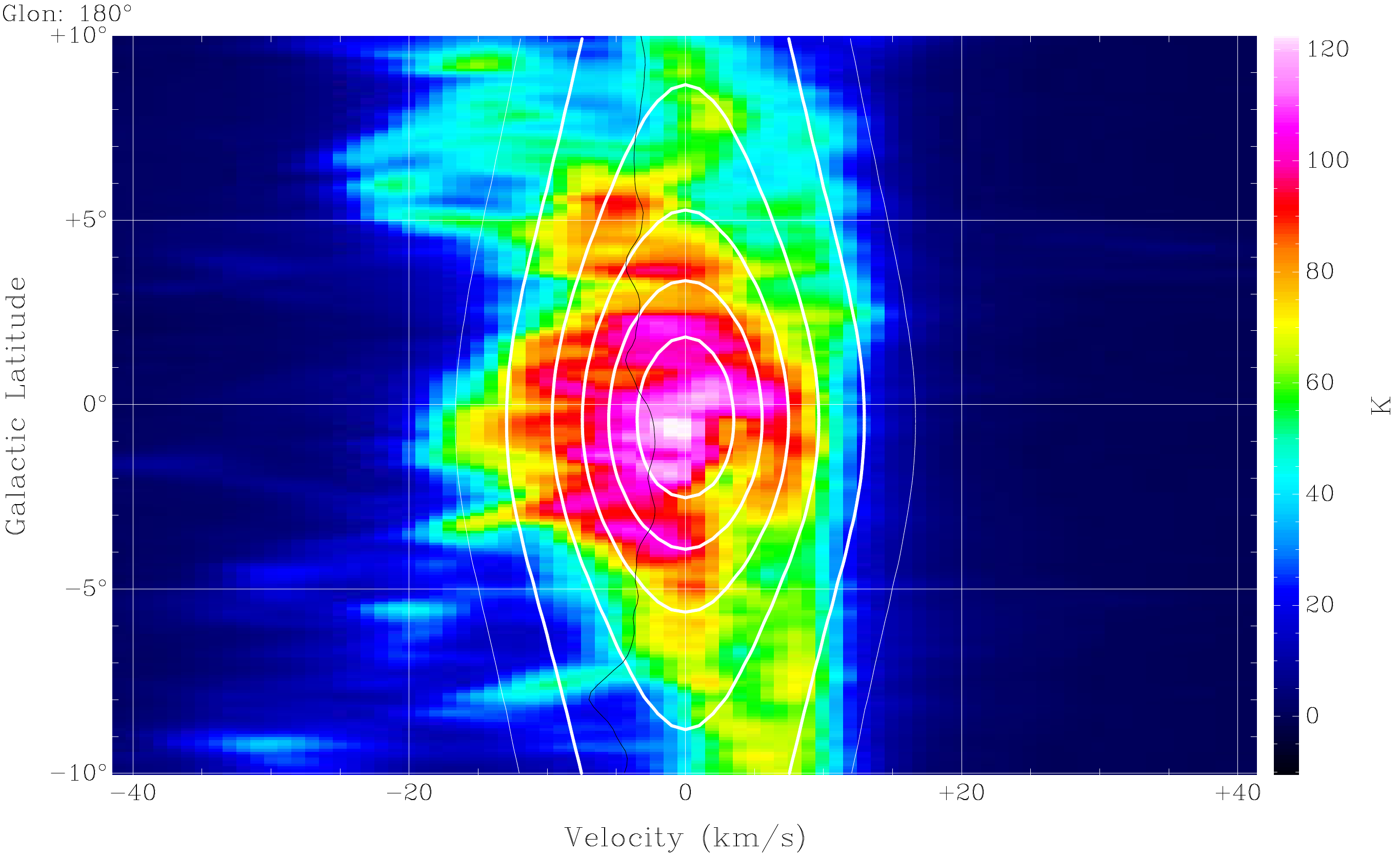}
\caption{Position--velocity diagram, color-coded for the HI emission in the
anticenter direction ($\ell =180^\circ $) derived from EBHIS observations. The isophotes at
10, 20, 40, 60, 80, and 100 K display the emission expected from a
Galactic mass model. The black line shows velocity centroids. Velocities
refer to the Galactic standard of rest, which at this particular longitude
is identical to the local standard of rest.}
\vspace{.2cm}
\label{Fig:HI_kalberla}
\end{figure*}

We extend our analysis using EBHIS observations for $\ell = 180^\circ$ and latitude $ -10^\circ < b < 10^\circ $. Figure \ref{Fig:HI_kalberla} shows a position-velocity
diagram for this case. The observed HI emission is shown color-coded and overplotted with model data. The model is based on a global fit of the
3D HI distribution in the Milky Way, including flaring and warping of the HI
disk (Kalberla et al. 2007 with updates from Kalberla \&
Dedes 2008 \footnote{Table 1 from Kalberla \& Dedes (2008) contains erroneous
densities for the cold and warm neutral media that are too high by a factor 1.4 due to
an inappropriate correction for He. Here we corrected this error.}). Clearly, from the equilibrium model of the
HI distribution no asymmetries in velocity are expected for all
latitudes at $ \ell = 180 ^\circ$. The observed HI emission shows, as usual, significant turbulent fluctuations in intensity and velocity that cannot be modeled. But for all latitudes there is a common trend that the emission is shifted to negative velocities. To quantify this velocity shift we determine velocity centroids 
($v_\mathrm{c}=\frac{\sum T v}{\sum T}$; Kerp et al. 2016), the first moment of the velocity
field, shown by the black line in Fig. \ref{Fig:HI_kalberla}. The average centroid velocity
for $|b| < 10\degr$ is -4 km/s, however, the HI emission for $|b| > 5^\circ $ is affected by
emission at negative velocities. For latitudes $b \ga 5^\circ $ there are
several clouds at $-25$ km/s$\la v_{\mathrm{LSR}} \la -10 $ km/s with peak
brightness temperatures $T_{\mathrm{B}} \sim$ 50 K. At $b \la 5^\circ $
this emission at intermediate negative velocities is weaker, 20 K$\la
T_{\mathrm{B}} \la 40$ K but more extended in velocity, -35 km/s$\la
v_{\mathrm{LSR}} \la -10 $ km/s. Emission at positive velocities
$v_{\mathrm{LSR}} \ga 15 $ km/s is nearly completely absent. Thus the
velocity centroid for $|b| > 5^\circ $ is biased toward negative
velocities. The observed emission at negative velocities is part of a
population of intermediate-velocity clouds, which are most likely located
in the transition region between the Galactic disk and halo (Stanimirovi{\'c} et al. 2006). 
Emission from similar clouds in the negative-velocity wings is insignificant close to the disk at 
$|b|<5^\circ$ (Fig. \ref{Fig:HI_kalberla}): we determine for
this range a centroid velocity of -3.5 km/s, which we consider
characteristic for the HI in the anticenter region.

Variations and systematic changes in the velocity dispersion of the HI
along the line of sight and uncertainties in the density
distribution imply uncertainties in the shape of calculated model
profiles in Figs. \ref{Fig:calc_HI} and \ref{Fig:HI_kalberla}, but they do not affect the velocity centroids. The center velocities of individual gas clouds along the
line of sight at $\ell=180\degr$ depend merely on the dynamics of the
system.

\subsection{Analysis in regions different from the anticenter}

The same analysis may be carried out in regions away from the anticenter, but in these cases
the systematic error is much larger because the uncertainties in the rotation curve have an
important effect on the conversion of heliocentric to Galactocentric coordinates.

\begin{figure*}
\vspace{1cm}
\centering
\includegraphics[width=14cm]{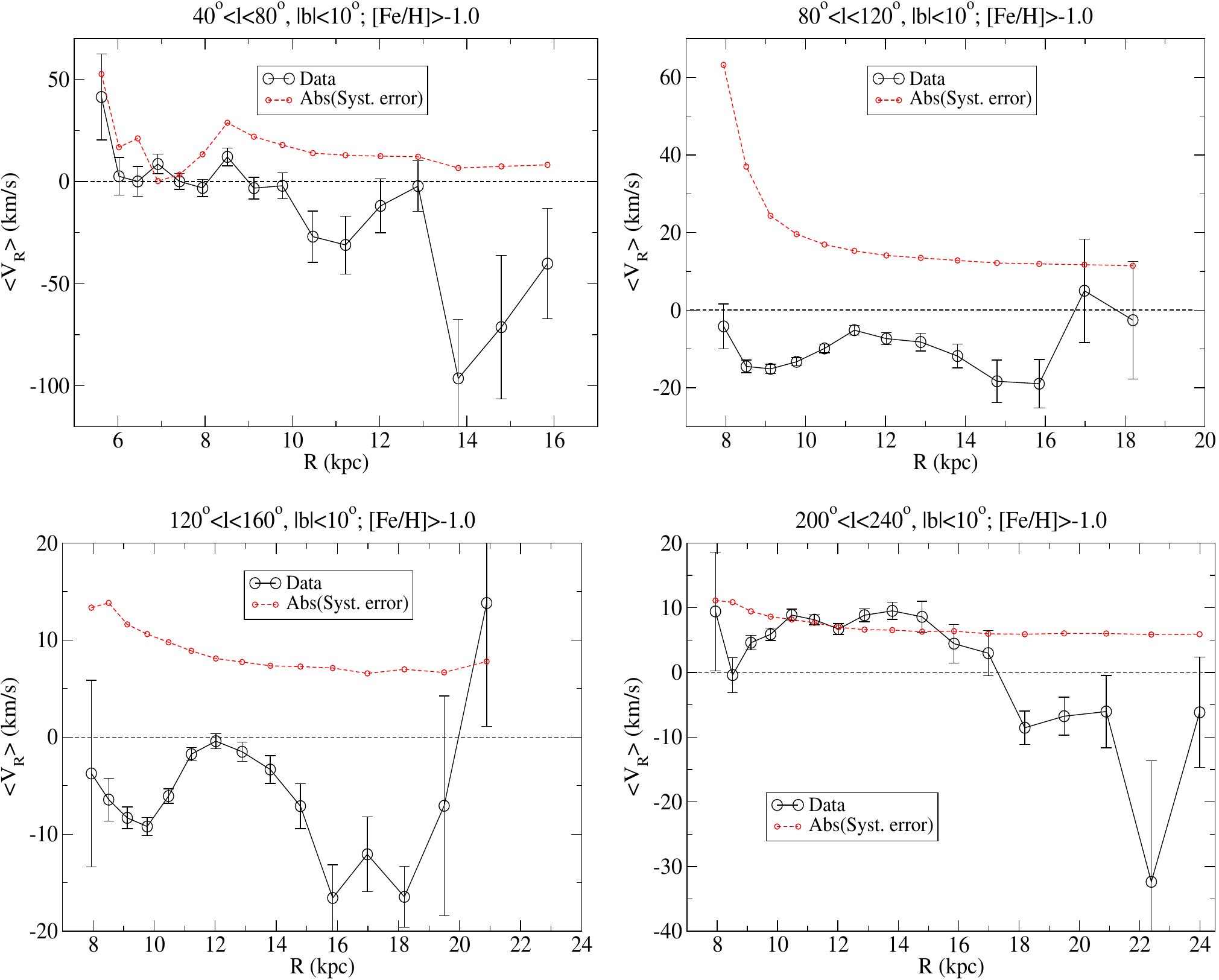}
\caption{Average radial Galactocentric velocity derived from Eq. (\protect{\ref{VR}}) with radial heliocentric velocities from APOGEE within regions different from the anticenter, including only stars with [Fe/H]$>$-1. Error bars represent statistical errors only. Systematic errors due to uncertainties in the parameters are plotted separately.}
\vspace{.2cm}
\label{Fig:noanticenter}
\end{figure*}

The results are plotted in Fig. \ref{Fig:noanticenter}, where we can see that the average
velocities are of the order of the systematic errors, rendering these measurements unuseful.
We need the proper motions to reduce these systematics in these directions, a work that was already
done with Gaia-DR2 data in G18 up to $R\approx 13$ kpc or LS18 up to $R\approx 20$ kpc. 

\section{Interpretation of results}
\label{.interp}

As shown by LG16, the detection of ${v}_{R}\ne 0$ in a certain sky region means that the average orbit of the stars is not perfectly circular, either because it is elliptical or because there is a component of secular expansion/contraction 
either of the whole disk or of a part of it.

\subsection{Elliptical orbits}

\begin{figure}
\vspace{1cm}
\centering
\includegraphics[width=8cm]{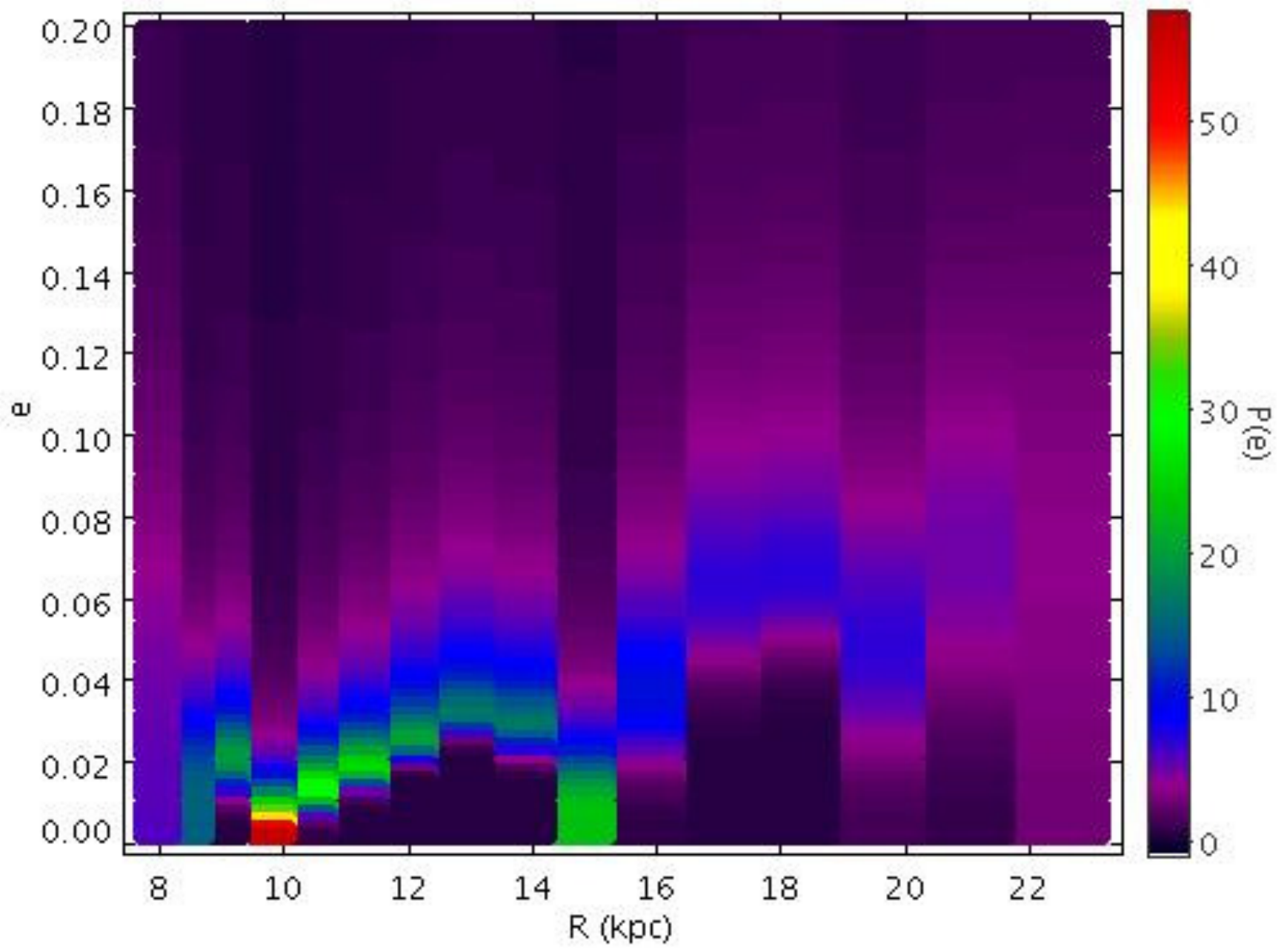}
\caption{Probability distribution of the eccentricity $e$ per unit $de$
given by Eq. (\ref{pe}) as a function of $R$, from $\langle v_R\rangle(R)$ derived
for the anticenter data with [Fe/H]$>-1$, assuming elliptical mean orbits.}
\vspace{.2cm}
\label{Fig:prob_ecc}
\end{figure}

In the case of elliptical orbits, ${v}_{R}\ne 0$ would indicate an eccentricity, $e$, different from zero. We have no information on the dependence of $v_R$ on $\phi $, so we cannot derive the exact orbits, but we can statistically derive 
the most likely values of $e$ for the mean orbit\footnote{The mean orbit is that followed by the average motion of large groups of stars in the same position, which is not the same as the orbit for each of the individual stars. For instance, the motion of the LSR has a closed orbit around the Galactic center, 
which is different from the motion of the Sun.}. 
In the approximation of low $e$, the probability of having a given value of
the eccentricity $e$ when we measure
a radial Galactocentric velocity \footnote{Do not confuse this with the average of the radial velocity over the whole orbit: here we only have $v_R$ for a given azimuth.}  $v_R$ with rms $\sigma$,  
   at a particular point of the orbit,
 is (LG16, Eq. (11)):
\begin{equation}
\label{pe}
P(e)de\approx \frac{2^{1/2}de}{\pi^{3/2}\sigma \,V_\phi \,e^2}
\int _{-V_\phi \,e}^{V_\phi \,e}
\frac{|x|}{\sqrt{1-\frac{x^2}{V_\phi ^2e^2}}}exp\left[-\frac{(x-v_R)^2}{2\sigma ^2}\right]dx
\end{equation}
where, as mentioned above,  
$V_\phi $ is the azimuthal velocity (as usual, we take the rotation curve data of
Sofue 2017, included in his Fig. 4, and we subtract the asymmetric drift as above; we neglect here the variation with $z$). 
Using this expression, we can evaluate the most likely values of the eccentricity as a function of $R$ derived from our results in the previous section. 
We carry out the calculation for the anticenter data with [Fe/H]$>-1$, without taking into account the systematic errors (we take $\sigma $ equal to the statistical error), 
which are plotted in Fig. \ref{Fig:prob_ecc}. We can observe that values of $e\approx 0.07$ are the most likely for $R\approx 18$ kpc, and lower values for other Galactocentric radii.

\subsection{Secular Expansion/Contraction in the Anticenter}

\begin{figure}
\vspace{1cm}
\centering
\includegraphics[width=8cm]{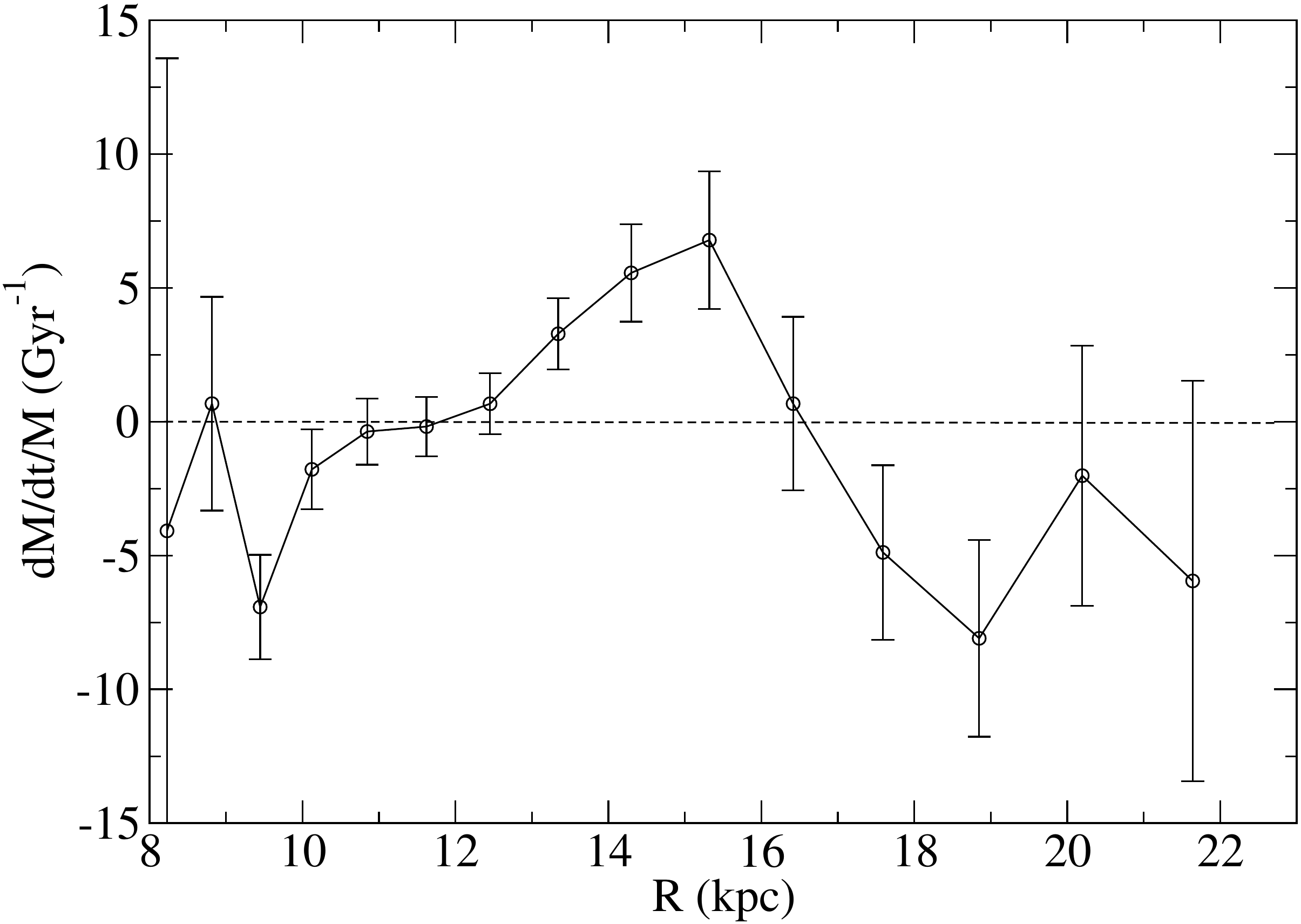}
\caption{Relative mass gain/loss given by Eq. (\ref{mdot}) from $\langle v_R\rangle(R)$ derived for the anticenter data with [Fe/H]$>-1$.}
\vspace{.2cm}
\label{Fig:masvar}
\end{figure}

For a particular line of sight in the anticenter,
the relative variation of stellar mass within radii between $R$ and $R+dR$ is 
(LG16, Eq. (13)):
\begin{equation}
\frac{\dot{M}}{M}(R)=\frac{-1}{R\sigma (R)}\frac{d[\langle v_R\rangle (R)R\Sigma (R)]}{dR}
,\end{equation} 
where $\Sigma (R)$ is the stellar surface density, 
assuming a constant average mass/luminosity ratio throughout the disk. 
With an exponential disk $\Sigma (R)\propto e^{-R/h_R}$, we obtain that the relative gain of stellar mass toward the anticenter at radius $R$ is
\begin{equation}
\label{mdot}
\frac{\dot{M}}{M}(R)=
\langle v_R\rangle(R)\left(\frac{1}{h_r}-\frac{1}{R}\right)-\frac{d\langle v_R\rangle (R)}{dR} \; .
\end{equation}
Assuming $h_R=2.0\pm 0.4$ kpc for a thin disk (L\'opez-Corredoira \& Molg\'o 2014), we get
the relative mass gain/loss given in Fig. \ref{Fig:masvar}.
It is indeed positive (gain) with a maximum amplitude at $R\approx 15$ kpc and negative (loss) with a maximum amplitude at $R\approx 19$ kpc.

The values of Fig. \ref{Fig:masvar} indicate a very fast evolution of the mass distribution. Most of the mass will be concentrated in the central regions without much change, but the outer disk has a trend at present to lose stars at $R\lesssim 11$ kpc and $R\gtrsim 16$ kpc and gain them at $R$ between 11 and 16 kpc. 
Assuming a surface density of stars in the solar neighbourhood of 27 M$_\odot $pc$^{-2}$ (McKee et al. 2015), the total variation of
stellar mass within $\Delta \phi =40^\circ =\frac{2\pi}{9}$ in our line of sight toward the anticenter
is: $-0.48\pm 0.13$ M$_\odot$/yr between R=9 and 11 kpc;
$+0.18\pm 0.06$ M$_\odot$/yr between R=11 and 16 kpc; $-0.03\pm 0.02$ M$_\odot$/yr between R=16 and 23 kpc. This is not possible if the Galaxy is in a  steady state, since the life of the
Galaxy is much longer than the time $\Delta T=\left (\frac{\dot{M}}{M} \right)^{-1}$ and the size of the Galaxy cannot change so fast.
This means that either there is a different trend for other azimuths,
   or
these velocities $v_R(R)$ are not
constant with time, so we live now in a period of fast expansion/contraction but this period will be short.
This mechanism may contribute to the variation of the disk size over cosmological times.  
Also some stars will escape the Galaxy and will be part of the halo once they abandon the disk. 

This net migration of stars might be related to the hypothesis known as
stellar radial migrations, which has indeed been useful to explain breaks in surface brightness (S\'anchez-Bl\'azquez et al. 2009), the formation of the thick disk (Sales et al. 2009), 
disk flaring (Vera-Ciro et al. 2014) or metallicity distributions (Casagrande et al. 2011; Grand et al. 2015) and age-metallicity distributions (Frankel et al. 2018). 

\subsection{Dynamical causes}

Minor mergers may raise vertical waves (G\'omez et al. 2013), but these perturbations
may not intensively affect the in-plane velocity (Tian et al.
2017). A local stream cannot be the explanation for radial velocities along a wide range of $\sim 20$ kpc, 
but it could be a large-scale stream associated with the Galaxy in the Sun--Galactic center line.
Since we do not have evidence of such a huge structure embedded in our Galaxy, we think this is not very likely. 
Moreover, there is a very small asymmetry between the northern and southern 
Galactic hemispheres (Fig. \ref{Fig:vr2}), thus we think it is very unlikely that the same stream independent of the Galaxy be so symmetric with respect to the plane.

In principle, the migrations of stars are theoretically expected as a consequence of the resonances of the bar and transient spiral arms in the disk (Sellwood \& Binney 2002; Roskar \& Debattista 2015; Halle et al. 2018; Hunt et al. 2018). No gradient in the radial velocity is expected from a bar effect for the observed distances (Monari et al. 2014), but it can be produced by spiral arms (Faure et al. 2014). 

\subsection{Spiral arms}

There might be accelerations due to spiral arms, which pull the stars toward the arm. In such a case, we would see fluctuations around zero in 
the positions where we cross a spiral arm: around $R=15$ kpc (the value at which $\langle v_R\rangle$ crosses the value of zero in Fig. \ref{Fig:anticenter}), and this is roughly
coincident with the position of the Outer (Norma--Cygnus) spiral arm (Camargo et 
al. 2015; Vall\'ee 2017). 
Therefore, the influence of the spiral arm looks like a plausible
candidate for driving our asymmetry. 
This mechanism, with compression where the stars enter the spiral arm and 
expansion where they exit, was also proposed as a tentative explanation for the
variations of more local radial velocities by Siebert et al. (2011), Monari et al. (2016) and G18.
The association of spiral structure potential with some ripples and ridges in the
kinematic data is also proposed (Faure et al. 2014; Hunt et al. 2018; Quillen et al. 2018) without the need to invoke an external perturbation, such as from the passage of the Sagittarius dwarf galaxy (Antoja et al. 2018).
In this case, one thing is remarkable from our data in Fig. \ref{Fig:anticenter}: the transition from the highest velocity to the lowest velocity is
quite smooth, over 4-5 kpc (although we must also bear in mind that the errors in the distance determination of the order of 10\% might also produce this smoothing), much larger than the 
typical observable width in spiral arms of $\sim 0.8$ kpc (Vall\'ee 2014).
Nonetheless, the transition from positive to negative velocities (or vice versa) is also smooth and short in 
the case of spiral arms and it is not directly linked to the inter-arm distance 
(Faure et al. 2014, Fig. 6). In the simulations by Faure et al., the value of $\langle v_R\rangle $ is expected 
to be quite stable in the inter-arm region.

Before we discuss this question, we want to comment
briefly on large-scale deviations from circular motions as observed in
the HI gas. The spiral density wave theory is based on the hypothesis of
quasi-stationary spiral structures (for a recent review see Shu
2016). Accordingly the motion of the gas is assumed to be forced by the
background gravitational potential of spiral arms. As the HI gas streams
through the spiral pattern it gets compressed and shocked, leading to
phase transitions from the warm to the cold neutral medium. Hydrodynamic
instabilities may cause the growth of knots in the spiral arms,
eventually leading to molecular clouds and star formation. Spiral
structure in the HI distribution is observed predominantly from velocity
crowding along the line of sight (Burton 1972).

These assumptions have been used by Roberts (1972) and Simonson (1976) (see 
also the review by Shu 2016), to model spiral structures in the gaseous Milky Way disk. From Figs. 5 and 6 of Roberts (1972), we estimate at $\ell = 180^\circ$ an expected
velocity shift of the HI line of -13 to -10 km/s for the Perseus arm,
approximately three times the centroid velocity measured by us. The
plots by Simonson (1976) show also some expected velocity shift, but it
is hard to read off numerical values. In both cases the Outer arm is not
part of the model. In any case, these numbers should depend on $R$.

Let us now consider this question: how much mass do we need in the spiral arm overdensity to produce the observed 
effects? In order to estimate it, let us assume a simple model in which a spiral arm is a ring with radius 
$R_{\rm sp}$, width $\Delta _{\rm sp}$ and total overdensity mass equal to $M_{\rm sp}$.
Of course, we know arms are spiral-shaped but for a simple calculation of the local
effect of the gravitational force this approximation is good enough.
The radial component of the acceleration of a star at Galactocentric radius $R$
is (Feng \& Gallo 2011)
\begin{equation}
\Delta a_R=2G\sigma \int _{R_{\rm sp}-\Delta _{\rm sp}/2}^{R_{\rm sp}+\Delta _{\rm sp}/2}
S\,\left[\frac{E(m)}{R(S-R)}-\frac{K(m)}{R(S+R)}\right]\,dS
,\end{equation}\[
m=\frac{4RS}{(R+S)^2};\ \sigma=\frac{M_{\rm sp}}{2\pi R_{\rm sp}\Delta _{\rm sp}}
,\]
$K(m)$ and $E(m)$ are respectively the first and second complete elliptical integrals.
The radial velocity of a star follows from  $v_Rdv_R=a_RdR$.
The distribution of mean velocities that we get in our Fig. \ref{Fig:anticenter} might be
a combination of particles with different initial conditions, some of them coming from the inner parts of the Galaxy with initial $v_R=0$ and others coming from the fall-off of particles from very high values of $R$. 
A simple estimation of the order of magnitude
integrating the previous motion equation numerically gives us that 
\begin{equation}
\mbox{Max}[v_R]\sim 100 \sqrt{\frac{M_{\rm sp}} 
{2\times 10^{11}\ {\rm M_\odot }} } \;\; \mbox {km/s}, 
\end{equation}
with a slight dependence on the width of
the spiral arms $\Delta _{\rm sp}$. If we compare this number with the value of
+6 km/s that we obtained in Fig. \ref{Fig:anticenter}, we get that the
mass of the whole spiral arm should of the order of $\sim 7\times 10^8$ M$_\odot $. 

If we have four spiral arms in our Galaxy with the same mass as this one, 
the total mass of the spiral arms would be $\sim 3\times 10^9$ M$_\odot $, which is
about 3\% of the Galactic disk mass (P\'erez-Villegas et al. 2015). This
is precisely the mass that is attributed to the spiral arms in the models
(P\'erez-Villegas et al. 2015), in agreement with other observations.

Kalberla et al. (2007) also posited the existence of a dark ring at 13 kpc$< R <$ 18.5 kpc,
in order to reproduce the observed flaring in the HI distribution; however, they attribute
a mass to this dark ring of $(2.2-2.8)\times 10^{10}$ M$_\odot $, which is much larger
than our estimation of the necessary mass of a ring.
The general assumption of Kalberla et al. (2007) was that the HI distribution is isothermal, which means velocity dispersions that cause the observed scale heights are constant with $R$; possibly this was a wrong assumption.
Anyway, apart from the calculation of the mass, apparently something is taking place at 
$R\approx 15$ kpc that produces important kinematical effects.


\subsection{Velocity Fields in out-of-equilibrium Structures: Radial Motions, Dispersion and Dynamical Equilibrium}
\label{.sylos}

In this section we discuss the kinematic properties of two different kinds of system that are caused by different dynamical processes and that can be compared, only from a qualitative point of view, with the observed velocity field. On the one hand we consider a mock galaxy catalog that was built from a cosmological $N$-body simulation, and on the other hand we discuss
the properties of some  systems created by the gravitational collapse of isolated overdensities. 
The signature of the dynamical processes at work in the two cases is imprinted in the amplitude of radial motions:
let us start from  the latter case.

In B17, it was shown  that the purely 
self-gravitating systems evolving from quite simple
initial configurations can  give rise easily
 to produce a quasi-planar spiral structure  
surrounding a virialized core and  qualitatively resembling a spiral galaxy. 
In particular,  for a broad range of non-spherical and non-uniform rotating
initial conditions, gravitational relaxation gives rise quite generically to a rich variety of structure 
characterized by spiral-like arms; the main characteristic of these structures  is that 
they are features of the intrinsically out-of-equilibrium nature of the system's collapse. 
That is, they represent long-lived non-stationary configurations 
characterized by predominantly radial motions in their outermost parts, but they also incorporate
an extended flattened region that rotates coherently about a well virialized core of triaxial shape
with an approximately isotropic velocity dispersion. Let us consider a few examples.

To characterize the 
velocity field, we consider the radial component of a particle's velocity $\vec{v}$,   
\begin{equation}
\label{vr}
v_r = \frac{ \vec{v}\cdot \vec{r}}{|\vec{r}|} 
\end{equation}
and the  vectorial ``transverse velocity" 
\begin{equation}
\label{vt}
\vec{v}_t (r)= \frac{\vec{r} \times \vec{v}(r)}{|\vec{r}|} \;,
\end{equation}
i.e., the vector of which the magnitude is that of the non-radial component
of the velocity, but oriented parallel to the particle's angular momentum 
relative to the origin. 
We will refer hereafter to the average of a quantity in a spherical shell with a
radius $r$: coherent rotation of all the
particles in the shell about the same axis then corresponds to
$\langle |\vec{v}_t| \rangle$= $|\langle \vec{v}_t \rangle|=v_\phi$.
The asymmetric drift is defined as (Binney \& Tremaine 2008),
\begin{equation}
\label{va}
v_a = v_c - \overline{v_\phi}
\end{equation}
where 
$v_c$ 
is the circular speed that corresponds to perfectly circular orbits, i.e.,
\begin{equation}
\label{vc} 
v_c^2(r)= \frac{GM(r)}{r} \;.
\end{equation}
To characterize the kinematics further, we also consider
the anisotropy parameter $\beta(r)$ 
\begin{equation}
\label{beta}
\beta(r)  = 1 - \frac{\langle v_t^2(r) \rangle}{2  \langle v_r^2(r) \rangle}  \;, 
\end{equation}
where  $\langle v_t^2(r) \rangle$ and $\langle v_r^2(r)\rangle$ 
are respectively the average square values of the transverse
and radial velocities.

\begin{figure}[htb]
\includegraphics[width = 3.5in]{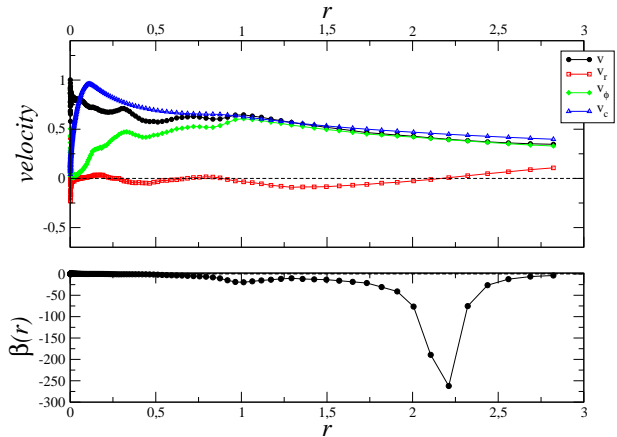} 
\caption{Upper panel:  velocity 
components measured in bins along the radial direction
for the simulation A1 at time $t=50$. 
$v$ is the total velocity, $v_r$ is the radial component (Eq.\ref{vr}),
$v_\phi$ is the circular component (see text for details), and
$v_c$ is the circular speed that corresponds to perfectly circular orbits 
(Eq.\ref{vc}).
Distance, velocity and time are given in units of the simulation’s constants $(r_0, v_0, r_0/v_0)$.
Bottom panel: anisotropy parameter $\beta(r)$ (see Eq.\ref{beta}).}
\label{D15-dyn} 
\end{figure}

\begin{figure}[htb]
\includegraphics[width = 3.5in]{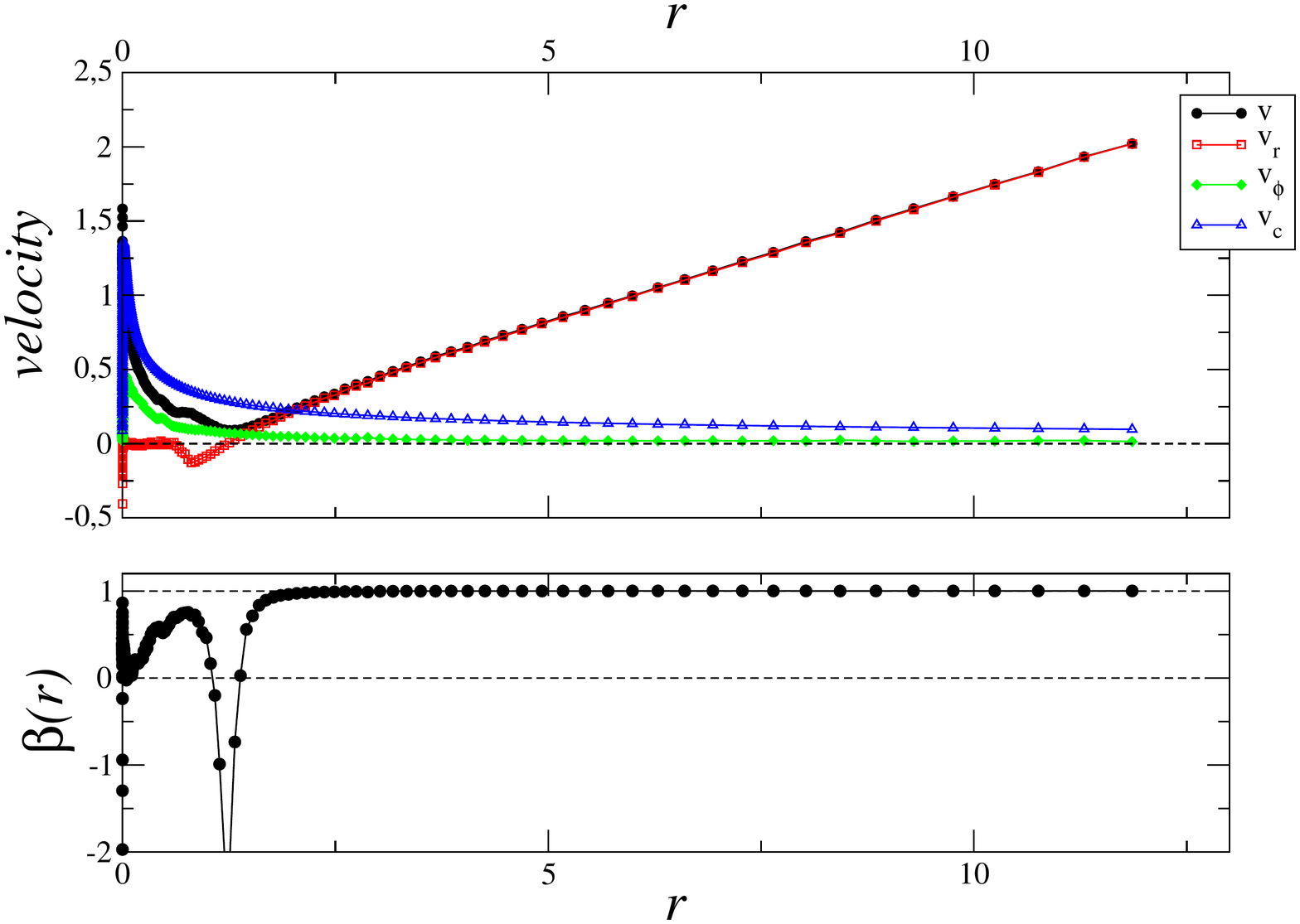} 
\caption{The same as Fig.\ref{D15-dyn} 
for the simulation A2 at time $t=100$. } 
\label{S6-dyn} 
\end{figure}

\begin{figure}[htb]
\includegraphics[width = 3.5in]{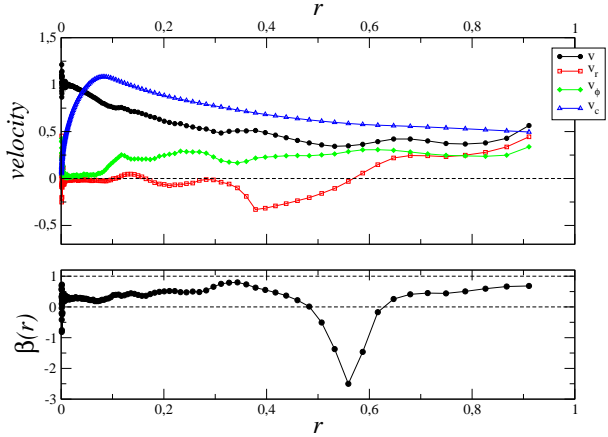} 
\caption{The same as Fig.\ref{D15-dyn} for the simulation A3 at time $t=50$.} 
\label{D12-dyn} 
\end{figure}

Let us now consider  three examples of this class of system
(see Figs. \ref{D15-dyn}-\ref{D12-dyn}): 
the initial 
conditions were  represented  by uniform and isolated ellipsoids 
(with different ratio between the semi-axes) and  
with normalized spin parameter (Peebles 1969) $\lambda=0.1-0.3$
(see B17 and Benhaiem et al 2018 for more details). 
We note  that at sufficiently small scales 
all simulations are characterized by a compact core with 
isotropic velocity dispersion, for which $\beta(r) \approx 0$. 
At larger distances from the system's center, 
the simulation A1 is characterized by a regime where 
rotational motions dominate, i.e. $v \approx v_\phi \approx v_c$ and   $\beta(r) \ll 0$.
In this case, the radial velocity remains small even in the outermost region of the system. 
On the other hand, the simulation A2, in an intermediate range of distances larger than the core's size
but with $r/r_0 \le 1$, is
dominated by rotational motions   
(i.e., $v_\phi  >0$) and has negligible net radial motions, 
i.e. $\langle v_r \rangle \approx 0$:
however, it shows a non-zero radial velocity dispersion so that 
$0 < \beta(r) <1$ and thus the asymmetric drift becomes 
large. Then at larger scales, 
the system is completely dominated by radial motion so that $\beta(r) \approx 1$.
Finally the simulation A3 shows an intermediate situation between 
A1 and A3: radial motions, in the very outermost region of the 
system, remain of the same order as the rotational ones. 

\begin{figure}[htb]
\begin{center}
\includegraphics[width = 3.0in]{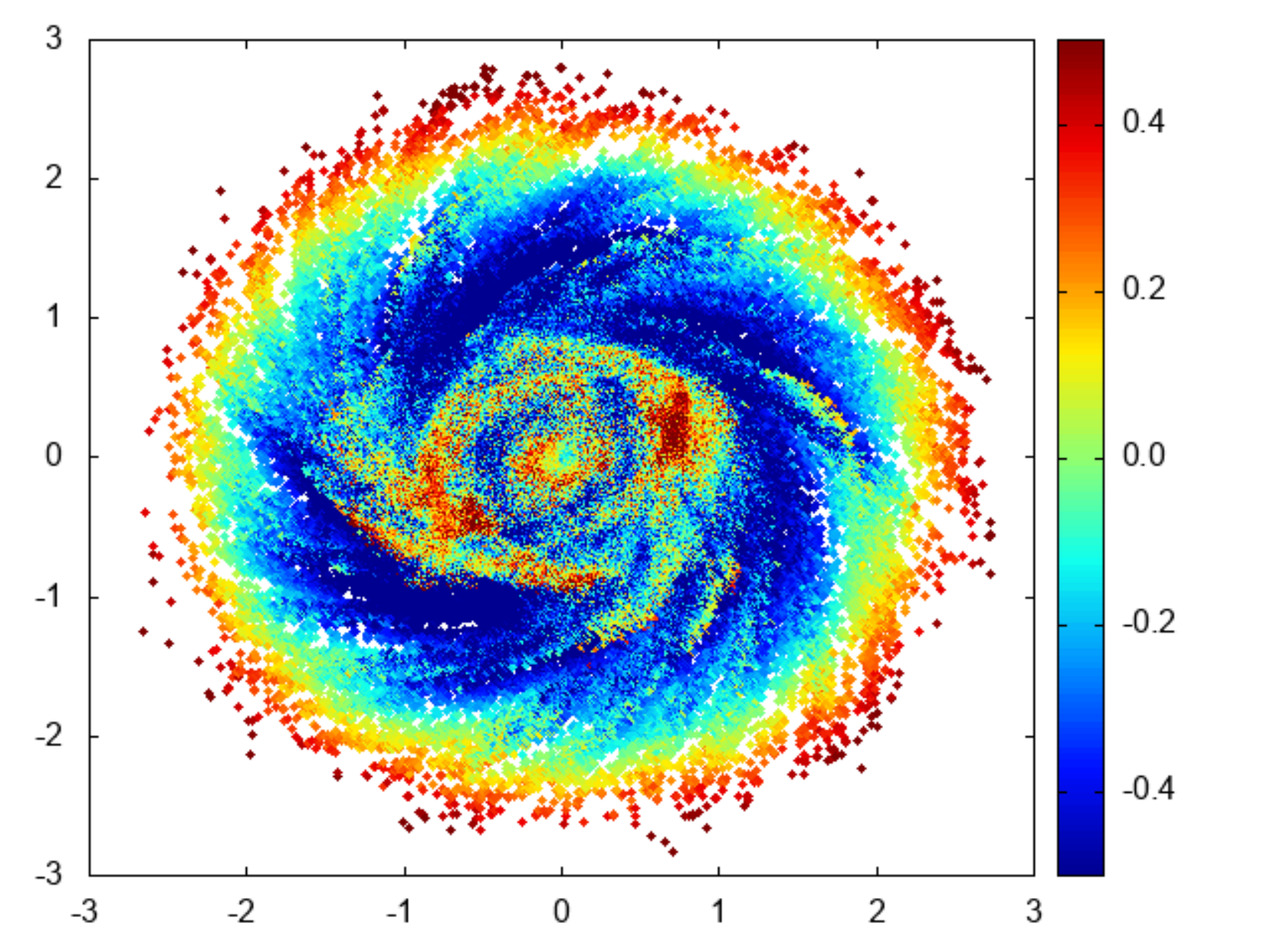} \\
\includegraphics[width = 3.0in]{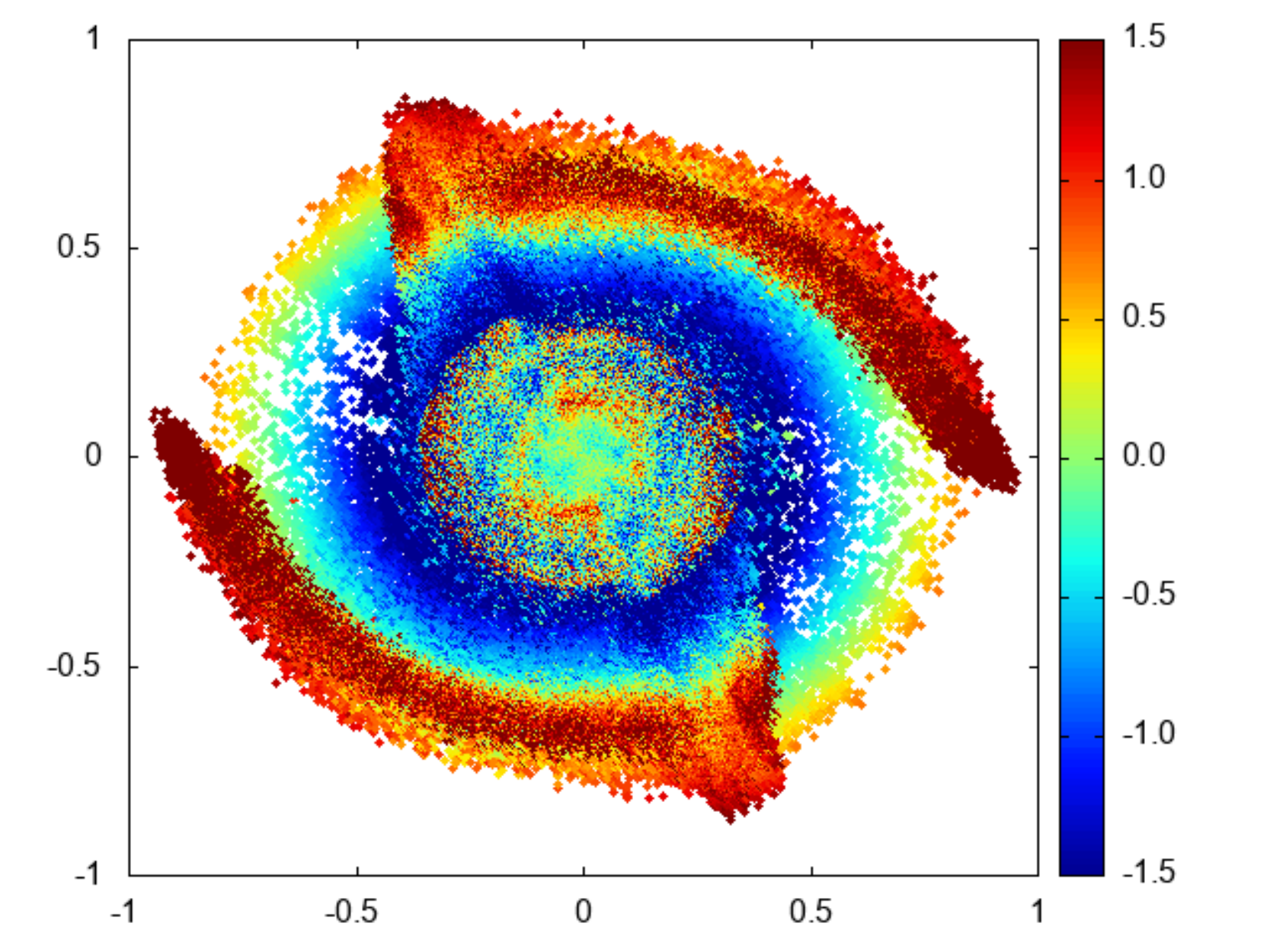} \\
\includegraphics[width = 3.0in]{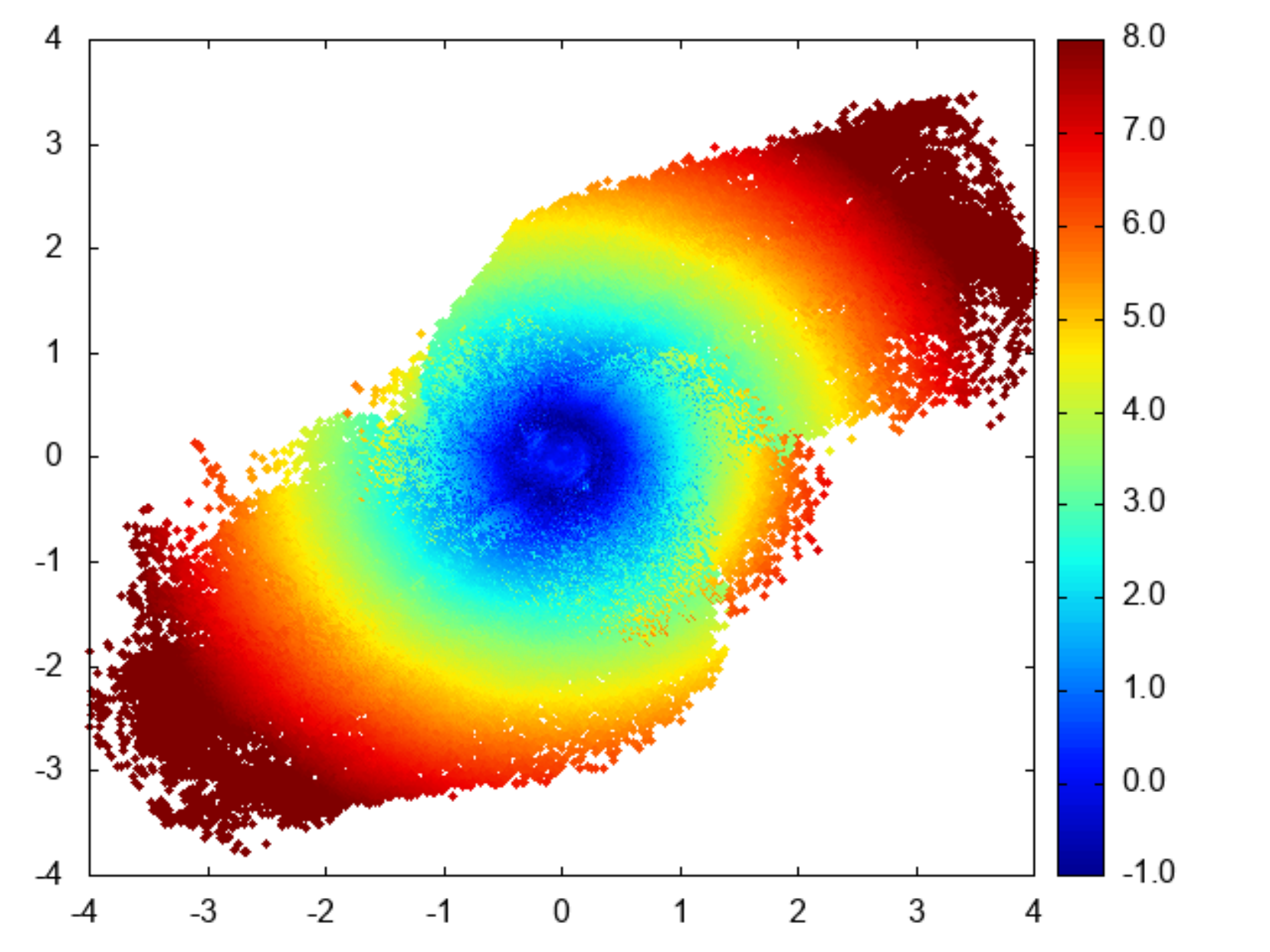} \\
\caption{Projection of the three simulations (from top to bottom: A1, A2, and A3)
on the plane identified by the two largest semi-axes. The color code 
 corresponds to the amplitude of the radial velocity. 
Distance and  velocity are given in units of
the simulation's constants $r_0,v_0$).
} 
\label{radialvel}
\end{center} 
\end{figure}

By a closer inspection we may note that, in all cases,  
the behavior of the average radial velocity, when different from zero,
 is first positive but close to zero, then negative and then positive again. 
This is due to the fact that there is a stream of particles,
spatially correlated, that are expanding 
outward and they produce an infall motion
of the shells that lie in front of them (see
Fig. \ref{radialvel}). The  amplitude 
of the negative radial velocities reduces with time.  Thus, a
 region where the  radial velocity is negative is thus a natural outcome of these out-of-equilibrium  models. 

{
The two cases where radial motions dominate in the outermost part of the systems, i.e. A2 and A3,
are characterized by a noticeable break of axisymmetry, while in the case where radial motions
are smaller (i.e, A1) the system remains close to axisymmetric. This is due to the fact that 
radial motions develop during the collapse because of the amplification of the initial
anisotropy: indeed, the simulation A1 was started from an oblate ellipsoid 
and thus remains symmetric in the plane of the two largest semi-axes --- while it contracts
in the orthogonal direction parallel to the minor semi-axis. 
Thus, because of the correlation between the direction of the radial motion and the direction 
of the major semi-axis, one must take into account that  a measurement
of the radial velocities in the direction of the anticenter of a given observer (in 
the real case, the Sun) is not generally expected to give a strong radial signal in these
simulations, unless such a direction is (by chance) aligned with the 
the direction of the largest semi-axis or if the departure from axisymmetry is small enough. 
On the other hand, the more the system is axisymmetric, the higher is 
the probability that a random observer sees at its anticenter a growing radial velocity 
but the smaller is its amplitude.


Finally,  
let us now briefly discuss the difference between the models 
that we have presented, where the formation of a disc-like flat structure is 
caused solely by a gravitational, and thus dissipationless, dynamics
and models in which instead the formation of a disc-like structure is 
driven by dissipational effects. 
Fig. \ref{Au6_vel}  shows the velocity components of a mock stellar catalog 
that was built from a cosmological simulation of the Auriga Project \footnote{Data for the 
simulation called Au6  are available at 
{\tt https://auriga.h-its.org/gaiamock.html}}. 
This  
 provides   a large suite of high-resolution magnetohydrodynamical 
simulations of galaxies simulated in a fully 
cosmological environment by means of the `zoom-in'  technique (Grand et al. 2018).
One may note in Fig. \ref{Au6_vel} that the average 
 radial component of the velocity is almost  zero at all distances, 
while  the velocity of stars is dominated by the circular component.
In addition, the radial velocity dispersion is of the same 
order as the tangential one 
 at large radii, while $\beta(r) < 0$ at smaller radii where the motion
 is predominantly circular.

\begin{figure}
\includegraphics[width = 3.5in]{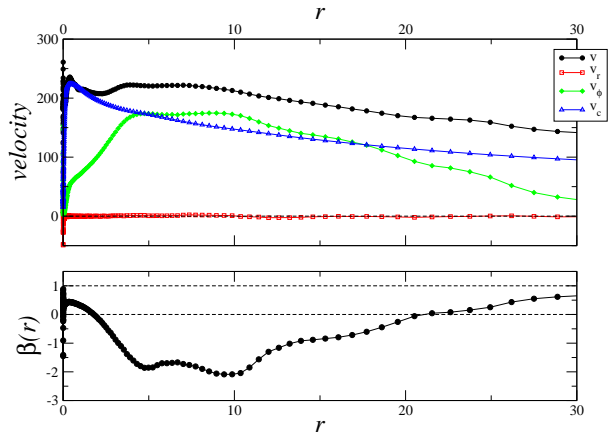}
\caption{The same as Fig.\ref{D15-dyn} 
for the Au6 simulation (stars).
Distance is in units of kpc and velocity in units of km/s.}
\label{Au6_vel} 
\end{figure}
{Fig. \ref{Au_proj}  shows that the radial velocity field is not characterized by any 
coherent spatial structure at large enough distances from the center (i.e., $r> 5$ kpc), 
where the system is very close to axisymmetric.
As mentioned above, the departure from axisymmetry is related to the presence of 
a complex velocity field, where radial motions become predominant at large distance:
systems of this kind can be generated by a dynamics that involves globally the whole
structure, i.e. a collapse, rather than a bottom-up aggregation process of the type
at work in cosmological simulations. 
}
\begin{figure}
\includegraphics[width = 3in]{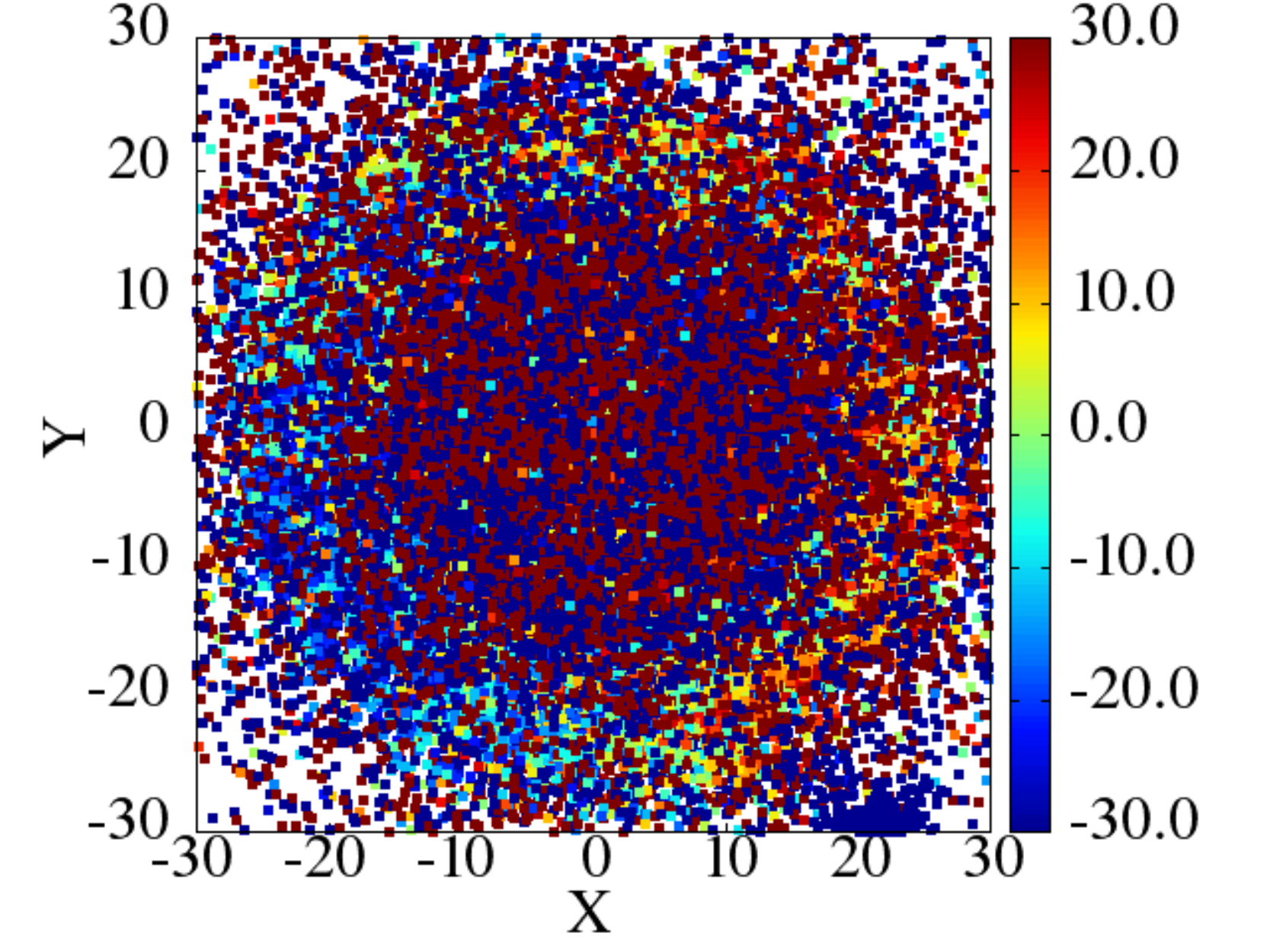}
\caption{As Fig.\ref{radialvel} but for the particles labeled as "stars" in the  Au6 simulation
that lie in the disk (i.e. -$2\; \mbox{kpc} \le Z \le 2 \; \mbox{kpc}$).
Distance is in units of kpc and velocity in units of km/s.}
\label{Au_proj} 
\end{figure}

\section{Conclusions}
\label{.concl}

Slightly non-circular mean stellar orbits in the Galactic disk were already known 
for $R\lesssim 16$ kpc (Siebert et al. 2011; Williams et al. 2013; LG16; Tian et al. 2017; Wang et al. 2018b; Carrillo et al. 2018; G18; LS18), but the analysis carried out in the present paper for the anticenter direction spans a wider range of distances, 8 kpc$<R<28$ kpc, and 
better characterizes the Galaxy's kinematics on large scales. 
The outcome is an increase in positive $v_R$ (expansion of the disk) from $R\approx 9$ kpc to $R\approx 13$ kpc, reaching a maximum of $\approx 6$ km/s, and a decrease outward, reaching $\approx -10$ kpc for $R>17$ kpc (negative means contraction of the disk). The analysis in other directions different from the anticenter is not possible, due to the large systematic errors resulting when we have only radial heliocentric velocities, \footnote{Nonetheless, when
accurate proper motions are available, it would be possible to obtain information for 
other lines of sight, like in G18 or LS18.} associated with uncertainties in the rotation curve, which are generated in the conversion between heliocentric and Galactocentric radial velocities.
This feature of higher negative velocities is also observed in 21 cm HI maps toward the anticenter, possibly dominated by local gas, thus corroborating for the gas the asymmetry of velocities that shows the departure of circularity in the mean orbits.

Following LG16, we have analyzed the results in terms of mean ellipticity of the orbits
or secular expansion, possibly associated with stellar migration deduced by other data sets. 
We discuss some possibilities about the dynamical causes that create these non-zero mean $v_R$. One possible explanation might be the gravitational attraction of a spiral arm. As a matter of fact, we see a change of regime from positive to negative velocities in the position where we cross the Outer spiral arm (Camargo et al. 2015) in the anticenter, $R\approx 15$ kpc, as would be expected if an
arm with total mass $\sim 7\times 10^8$ M$_\odot $ (in agreement with the expected
mass of a spiral arm) pushed the stars with $R<15$ kpc (thus, $v_R>0$) and pulled the stars
with $R>15$ kpc (thus, $v_R<0$).

An unconventional theoretical model was explored using the B17 dynamical model:
a simple class of out-of-equilibrium, rotating, and asymmetrical mass distributions that
evolve under their own gravity to produce a quasi-planar spiral structure  surrounding a virialized core. 
Non-circular orbits in the outer disk are precisely one of the
predictions of this model. This scenario is also related to spiral arms and their formation, but it is not the direct gravitational attraction of the spiral arms that produces the mean radial velocities. Rather, it is the fact that orbits in the very outer disk are out of equilibrium so they have not reached circularity yet.
Under some initial conditions, our simulations with the B17 model are able to reproduce the observed feature of $v_R$ in Fig. \ref{Fig:anticenter} and predict that at $R$ much larger than 30 kpc we should return again to the regime of positive $v_R$ with very large amplitude (of the order of the rotation speed). This is something that we cannot check with the present data, but it would be desirable to do in the future.

Certainly, further kinematic information at farther distances or along different lines of sight might constrain better the dynamical scenarios of our Galaxy. G18 already provided some $v_R(\phi )$ analysis, but constrained by $R<13$ kpc, and longer distances need to be explored. The extension of the kinematical maps with Gaia-DR2 made by LS18 gives a better insight to farther distances, and the future releases of Gaia data will extend those maps
even more.
And also in other galaxies, 2D spectroscopy [for instance, or radio data of THINGS (Sylos Labini et al. 2018), or using Multi Unit Spectroscopic Explore (MUSE) at VLT telescope] may allow us to carry out an analysis of non-circularity in mean orbits.
At present, at least,  we can say that non-zero radial velocities in the outer disk are a fact, and this should persuade us to abandon the general idea that spiral galaxies are in a quasi-equilibrium configuration where stars move on steady circular orbits around the center of the galaxy.

\begin{acknowledgements}
Thanks are given to the anonymous referee for helpful comments and 
to S\'ebastien Comer\'on for helpful comments on a draft of this paper.
M.L.C. was supported by the grant AYA2015-66506-P of the Spanish Ministry of Economy and Competitiveness (MINECO). 
F.S.L. was granted access to the HPC resources of The Institute for
scientific Computing and Simulation financed by Region Ile de France
and the project Equip@Meso (reference ANR-10-EQPX-29-01) overseen by
the French National Research Agency (ANR) as part of the
Investissements d'Avenir program. 
Funding for the Sloan Digital Sky Survey IV has been provided by the Alfred P. Sloan Foundation, the U.S. Department of Energy Office of Science, and the Participating Institutions. SDSS-IV acknowledges
support and resources from the Center for High-Performance Computing at
the University of Utah. The SDSS website is www.sdss.org.
SDSS-IV is managed by the Astrophysical Research Consortium for the 
Participating Institutions of the SDSS Collaboration including the 
Brazilian Participation Group, the Carnegie Institution for Science, 
Carnegie Mellon University, the Chilean Participation Group, the French Participation Group, Harvard-Smithsonian Center for Astrophysics, 
Instituto de Astrof\'isica de Canarias, The Johns Hopkins University, 
Kavli Institute for the Physics and Mathematics of the Universe (IPMU) / 
University of Tokyo, Lawrence Berkeley National Laboratory, 
Leibniz Institut f\"ur Astrophysik Potsdam (AIP),  
Max-Planck-Institut f\"ur Astronomie (MPIA Heidelberg), 
Max-Planck-Institut f\"ur Astrophysik (MPA Garching), 
Max-Planck-Institut f\"ur Extraterrestrische Physik (MPE), 
National Astronomical Observatories of China, New Mexico State University, 
New York University, University of Notre Dame, 
Observat\'ario Nacional / MCTI, The Ohio State University, 
Pennsylvania State University, Shanghai Astronomical Observatory, 
United Kingdom Participation Group,
Universidad Nacional Aut\'onoma de M\'exico, University of Arizona, 
University of Colorado Boulder, University of Oxford, University of Portsmouth, 
University of Utah, University of Virginia, University of Washington, University of Wisconsin, 
Vanderbilt University, and Yale University.

\end{acknowledgements}

\end{document}